# Bright focused ion beam sources based on laser-cooled atoms

J. J. McClelland,[1] A. V. Steele,[1,2] B. Knuffman,[1,2] K. A. Twedt,[1,3]* A. Schwarzkopf,[1,2] and T. M. Wilson[1]*

[1]Center for Nanoscale Science and Technology

National Institute of Standards and Technology

Gaithersburg, MD 20899

[2]zeroK NanoTech, Gaithersburg, MD 20878

[3]Maryland Nanocenter, University of Maryland, College Park, MD 20742

## Abstract

Nanoscale focused ion beams (FIBs) represent one of the most useful tools in nanotechnology, enabling nanofabrication via milling and gas-assisted deposition, microscopy and microanalysis, and selective, spatially resolved doping of materials.  Recently, a new type of FIB source has emerged, which uses ionization of laser cooled neutral atoms to produce the ion beam.  The extremely cold temperatures attainable with laser cooling (in the range of 100 µK or below) result in a beam of ions with a very small transverse velocity distribution. This corresponds to a source with extremely high brightness that rivals or may even exceed the brightness of the industry standard $Ga^+$ liquid metal ion source.   In this review we discuss the context of ion beam technology in which these new ion sources can play a role, their principles of operation, and some examples of recent demonstrations.  The field is relatively new, so only a few applications have been demonstrated, most notably low energy ion microscopy with Li ions. Nevertheless, a number of promising new approaches have been proposed and/or demonstrated, suggesting that a rapid evolution of this type of source is likely in the near future.

**\*present address: Science Systems and Applications Inc., Lanham, MD 20706

## I.  Introduction

Nanotechnology has witnessed extraordinary growth in recent years, blossoming into a far-reaching, cross-disciplinary enterprise that touches such diverse fields as physics, chemistry, materials science, electrical engineering and biology.  Fueling this growth has been a steady development of new instrumentation enabling measurement and fabrication on the nanoscale – tasks that are not at all trivial when one considers the challenging requirements for spatial precision and accuracy, not to mention stability and detection sensitivity, that arise when one is working on the scale of just a few atomic spacings.  Such measurement developments as electron microscopy and scanning probe microscopy, coupled with remarkable advances in nanofabrication via various forms of lithography and self-assembly, have proven to be critical enablers of the field.  But, like any technology, these innovations have their limitations, and it is clear that if nanotechnology is going to realize its full potential, further development of instrumentation is going to be critical.

One of the instrumentation developments that has come to play a central role in nanotechnology is the nanoscale focused ion beam (FIB).  In a configuration similar to a scanning electron microscope, it is possible to create a beam of ions focused to a spot size on the order of just a few nanometers.  Such beams have proven exceptionally useful in a wide range of applications, both in nanoscale measurement and in nanofabrication, and have found their way into a number of commercial instruments aimed at providing diverse functionality across several disciplines.

Key to the ability to create a nanoscale FIB is the ion source, which must provide enough ions with small enough spatial and angular spread to enable tight focusing.  Beginning in the late 1970s, ion sources with the necessary characteristics began to appear, and one type that soon came to dominate was the liquid metal ion source (LMIS).[1]  With an extremely simple construction, high brightness, and robust performance, the LMIS has become a central driver in the widespread adaptation of focused ion beam techniques in nanotechnology.

Other sources with the requisite characteristics have of course been under development, but it has not been until recently that the field has seen some significant advances.  One cause for advancement has been the perfection of the gas field ionization source (GFIS), a concept which actually predates the LMIS,[2] but has taken several decades to bring to the point of real practical application.  A larger reason, however, lies in the spread of nanotechnology and the corresponding drive to broaden source operation to encompass a wider range of ion species, a larger range of beam energies, and, of course, a smaller beam focal spot size.  This push has led to implementation of the GFIS with both He[3] and Ne,[4] extension of the LMIS to other species through use of liquid metal alloys,[5,6] and also to improved plasma-based sources, which can now make high-current, bright ion beams with a number of gaseous species, in particular heavy noble gases such as Xe.[7]  Driven by the needs of nanotechnology, all of these sources have advanced to a stage of commercialization, and are currently available in focused ion beam systems provided by a number of vendors.

It is in this context of a desire for better FIB performance that recent research has led to the development of the subject of this review, namely ion sources based on photoionization of laser-cooled atoms. As we will discuss below, this new type of source addresses the problem of generating a bright ion beam with an approach that is very different from the one taken by the LMIS and the GFIS. Instead of making the source size extremely small, the new source achieves high brightness by making the angular spread as small as possible. From the point of view of beam brightness, this approach can be considered ion-optically equivalent to starting with a small source size, and the result can be an ion source with brightness in the same range as the LMIS, or perhaps even higher, with a number of desirable characteristics: (a) potential access to over 27 new ionic species; (b) a very small energy spread, making possible a wide range of beam energies, in particular low energies; (c) a relative insensitivity to source vibrations; and (d) an inherent long-term stability with no need to rely on careful preparation of a sensitive tip.

In comparison with conventional ion sources, cold atom ions sources are in a very early stage of development. Nevertheless, some promising advances have appeared recently in the literature, from the very first publication of a proposed source,[8] to the demonstration of a fully operation cold-atom based lithium ion microscope.[9] Based on the work that has been done so far, it seems likely that opportunities for progress in this field will expand significantly in the near future. For this reason it is useful to review the present status with the intent of laying the groundwork for future developments.

This review is structured as follows: we begin by discussing a few applications of focused ion beams to set the context for the new source technology. Next we provide a brief review of the underlying principles that govern the operation of focused ion beam sources, where we introduce the concepts of beam emittance and brightness and discuss their importance for creating a FIB with a small focal spot. This section is followed by a brief review of conventional FIB sources. We then present a short discussion of laser cooling as it pertains to cooling the atoms used in a cold atom ion source. Following this, we discuss realizations of ion sources based on laser cooled atoms, in particular magneto-optical trap-based ion sources and cold atomic beam ion sources, and finish with some comments on the future outlook for sources of this type.

While the intention of this review is to be comprehensive, we have chosen not to cover the closely related use of photoionized laser-cooled atoms as an electron source. There has been significant research in this area as well, and a number of publications have shown that these sources have promise for producing highly coherent, fast bunches of electrons[10,11,12,13,14,15] for applications such as time-resolved electron diffraction of large biomolecules. Though the principles and operation of these electron sources are quite similar to the cold ion sources discussed here, the applications are generally quite different, so we will leave this subject for a possible future review.

## II. Applications of focused ion beams

The application of FIBs in nanotechnology has evolved into an extraordinarily broad and diverse discipline, and has been covered by a large number of reviews[16,17,18,19,20,21,22,23,24,25] and books.[26,27,28,29,30] For the purpose of the current review, we touch on a few of the more prominent applications in order to set the stage for ways in which the new sources discussed here can complement existing sources and provide enhanced performance in a number of instances.

*Nanofabrication*

Perhaps the most widely practiced application of focused ion beams is in nanofabrication – that is, the creation of nanoscale structures by removing and/or adding material to sculpt a desired form. Because FIBs can be focused to a probe size of just a few nanometers, it is possible to create structures with very high spatial resolution. Using such a focused ion beam, fabrication can proceed in a number of ways.

The simplest form of FIB nanofabrication is referred to as milling or micromachining, and corresponds to using the impact force of the incident ions to eject, or sputter, material from the target.[31] Depending on ionic species and ion beam energy, the sputter yield can be as high as five ejected atoms for each incident ion, resulting in a removal rate that exceeds the incident ion current in some cases. Beam currents on the order of picoamperes to nanoamperes are sufficient to fabricate nanostructures in a relatively short time. While micromachining is frequently used to directly fabricate objects on the nanoscale, it also has found extreme utility in cross-sectioning samples for scanning electron microscopy (SEM) and transmission electron microscopy (TEM).[32] Typically a FIB is used to carve out a trench in the sample of interest and then a SEM image of the sidewall is acquired, showing the buried structure of the sample. For a TEM sample, trenches are created on both sides of a thin lamella, which is then cut free with the FIB, removed, and placed on a TEM grid.[33]

Besides high spatial resolution, the most desirable characteristics of a FIB for micromachining are a high sputter yield, which favors heavy ionic species, and an ability to control surface roughness, which is facilitated by having some control over the ion beam energy. Also, it is sometimes important to avoid "staining" of the sample, which occurs when ions from the FIB are implanted in the target. Staining can cause artifacts in the imaging of cross sections, and also can "poison" the functionality of a nanostructure by introducing contamination. A choice of ionic species in the FIB can help alleviate this problem.

In addition to straightforward milling, it is possible to enhance the removal of material,[34] or even cause the deposition of additional material[35] by introducing a chemically reactive gas into the region where the ion beam strikes the target. This process has seen great success in a number of arenas, most notably in the areas of mask repair for the microelectronics industry,[36] and editing of integrated circuits, where the wiring in prototype chips can be exposed and rewired via ion-beam assisted etching and deposition.[37] While these processes will work with almost any ionic species, and in fact, also with electrons, they are optimized by choosing a species and beam energy that result in the highest chemical efficiency along with the least amount of surface damage and staining, with the additional consideration that removal of redeposited material through sputtering is sometimes beneficial. Also, of particular interest to circuit edit applications is attaining as small a probe size as possible, as technology pushes into the single-digit-nanometer regime.

Another form of nanofabrication for which FIBs have proven useful involves the direct implantation of ions to create nanoscale patterns of functionalized material. Diverse applications such as creation of semiconductor devices,[5] fabrication of quantum wells[38] or magnetic structures,[39] control of quantum dot formation,[40] or tailoring of surface plasmon-polaritons[41] have been realized with this method. In this type of application, it is perhaps even more important to have a choice of ionic species in the beam, because the implanted ion can play a functional role in altering the material properties. It is also important to be able to control implantation energy, because it governs the depth at which ions are implanted and the amount of damage created in the surrounding material.

It is also possible to use a FIB to create patterns in conventional resist-based lithography process.[42,43,44,45] In this case the sample is coated with a thin resist which is then exposed in a pattern with a FIB. During exposure the resist is chemically altered by the energy and charge of the incident ions, becoming either susceptible (positive resist) or immune (negative resist) to dissolution by a developer. After development, further etching transfers the pattern into the sample. Ion beam lithography can have some advantages over electron beam lithography, for example, (a) resists can be more sensitive to ions, (b) ions have a shorter penetration depth and smaller lateral straggle, and (c) ions have a very small de Broglie wavelength and hence effectively do not have a diffraction limit. While lithography does not depend critically on the ionic species, it still can be optimized by adjusting the resist sensitivity and penetration depth of the ions through species and energy choice, and it of course benefits from a FIB system with as high a current as possible in as small a probe size as possible.

*Ion microscopy*

While FIBs may find their most common application in nanoscale patterning or otherwise modifying a surface, they also have been shown to be quite useful as an imaging tool for microscopy.[3,46,47,48,49,50] For this application, the ion beam is used in much the same way as an electron beam is used in a SEM: a focused beam is rastered across the sample while backscattered particles are detected in synchrony with the scan to form an image.

Using ions instead of electrons introduces a number of significant differences in the imaging process, which can produce new surface information that is complementary to SEM data. The detected particles can be secondary electrons, in which case the image shows topography and surface work function variations as it does with secondary detection in the SEM, with some additional benefits: (a) the secondary yield tends to be higher per incident ion, (b) the ions straggle much less than electrons in the material, reducing spurious signals from outside the probe diameter, and (c) insulating regions of the sample charge positively and therefore trap any secondary electron emission, resulting in strong charge-based contrast with insulating regions appearing black. Alternatively, backscattered ions can be detected, in which case the detected signal is a strong function of the backscatter probability. Depending on the ion species being used, this can result in strong contrast based on the atomic mass of the target, allowing compositional differentiation. An additional benefit of detecting backscattered ions is that the signal can be much less sensitive to sample charging, since the scattered ions generally have an energy close to the beam energy, unlike secondary electrons whose energy is typically just a few electron volts. It is worth noting, however, that ion backscatter yield drops significantly as ion energy is increased, and so is only really practical if a suitable beam can be made with energy in the few-kiloelectronvolt range.

While imaging with ions has a number of benefits, there are also a few drawbacks that should be considered. The primary drawback is that ions can be more destructive than electrons, and if heavy ions are used at high impact energies, it is quite possible to destroy the object of interest by sputtering before enough signal has been acquired to form an image. This phenomenon can in fact lead to a fundamental limitation on the resolution attainable with a focused ion beam.[51] Another consideration is the fact that ions will generally be implanted in the sample during imaging, and this may not be desirable. Other drawbacks relate more to the state of the art of FIB sources than to the inherent nature of ion beam imaging. Today's sources do not typically provide quite the resolution attainable with a good field emission SEM, and the best resolution is attained only at high beam energies of several tens of kiloelectronvolts. Furthermore, only a few ionic species, in particular Ga[46,47,48] and He,[3,49,50] have been implemented with microscopy as the primary application. Having a wider choice of beam energies and ionic species would certainly allow ion microscopy to reach a much broader range of applications.

*Microanalysis*

In the spirit of microscopy, but with greater functionality, FIBs have additional capabilities that make them an especially valuable tool for nanotechnology. In a process known as secondary ion mass spectrometry (SIMS), ions can be used to purposefully sputter and ionize material from a target, and a mass spectrometer can then be used to analyze the ejected species.[52] This technique has proven to be extraordinarily sensitive and can be used to detect very small quantities of material with isotopic sensitivity. While most commonly used with only moderate spatial resolution, SIMS has been demonstrated with resolution as good as 50 nm using a gallium focused ion beam,[53,54,55,56,57] and, more recently, a cesium focused ion beam.[58,59] With the high spatial resolution provided by high quality focused ion beams, SIMS is proving to be an invaluable tool in a number of disciplines ranging from cell biology to planetary science. However, there is need for improved performance at these high resolutions, and much could be gained from improved FIB sources with higher brightness and greater choice in ionic species and beam energy.

Analysis of backscattered ions can also provide compositional information on the target. In a measurement referred to as Rutherford backscattering spectrometry, ions with relatively high energy are directed at a surface and the resulting backscattered ions are detected. Because of recoil of the target atoms, kinematics dictates that backscattered ions will experience some energy loss. This loss depends on the relative mass of the scattered ion and target atom, and on the scattering angle. As a result, the energy spectrum of the backscattered ions, collected in a narrow angular region, can be used to determine the mass of the target atoms. While Rutherford backscattering spectroscopy is typically done with high energy alpha particles having energies up to 3 MeV, and generally does not have resolution better than a few hundred micrometers, it has also been successfully implemented in GFIS-based He ion microscope to achieve nanoscale-resolution compositional mapping.[49]

A related analytical technique that has yet to be exploited fully with high resolution focused ion beams is low energy ion scattering (LEIS).[60,61,62] Because low energy ions scatter mostly from the surface of a material, LEIS is particularly suited to elemental analysis of surfaces in ultrahigh vacuum, where it is quite useful for studying composition and structure in surface chemistry, and can be uniquely sensitive to the top-most atomic layer of a sample. Spatial resolution has been modest, with reported resolutions as small as 5 μm.[63]  However, with improvement of focused ion beam sources, especially with high performance at lower energies, there is potential for better resolution and the possibility of a new, high-sensitivity nanoscale microanalysis tool.

## III. Focused ion beam source principles

At the most basic level, the requirements of a source for focused ion beams are relatively simple to state: deliver a beam of ions that can provide as much current as possible into as small a focal spot as possible, with an ionic species and a beam energy that are useful for one or more applications.  While these requirements are simply stated, the evaluation of the suitability of one source or another for a particular focused beam application is often more complicated and nuanced, and depends not only on the source but also the entire focusing system, as well as the specific requirements of the application.  That being said, there are some basic principles that can be used to characterize a source with regard to how well it should be able to perform in a focused ion beam application.

*Emittance*

A great deal of understanding of source performance can be had by considering an ion beam as a statistical ensemble of ions being accelerated, decelerated, focused, or drifting along the beam axis.  In this picture the motion of the ions in the beam can be characterized by an ensemble of trajectories, each with transverse position $(x, y)$ and transverse angles $(\alpha_x, \alpha_y)$, which vary with position $z$ along the beam.  With the assumption that the beam is not too strongly divergent or convergent, the angle $\alpha_x$ (or $\alpha_y$) can be related to the corresponding transverse momentum $p_x$ via $\alpha_x \approx v_x/v_z = p_x/\sqrt{2\,m\,U}$ , where $v_x$ is the transverse velocity, $v_z$ is the longitudinal velocity, and $m$ and $U$ are the mass and kinetic energy of the ion, respectively.  Using this relation it is possible to make a connection to Liouville's theorem,[64] which states that for an ensemble of particles moving in a conservative potential the volume in phase (i.e., position-momentum) space is conserved.  Invoking this conservation, we can define a conserved quantity $\varepsilon_x$, referred to as the normalized emittance, as

$$\varepsilon_x = \sqrt{m/2}(<x^2><v_x^2> - <xv_x>^2)^{1/2} \qquad (1)$$

where angle brackets indicate averaging over the ensemble.  With the approximation $\alpha_x \approx v_x/v_z$, we arrive at an expression for the emittance of an ensemble of particles travelling in a beam in the $z$-direction[65]

$$\varepsilon_x = (<x^2><\alpha_x^2> - <x\alpha_x>^2)^{1/2}\sqrt{U}. \qquad (2)$$

By examining eq. (2) we see that the first term in $\varepsilon_x$ represents the product of the spatial and angular spreads in the beam, the second term represents correlations between transverse position and angle, and the overall expression represents a measure of the area occupied by the trajectories in phase space.

It should be noted that eq. (2) is properly referred to as the normalized root-mean-square (RMS) emittance, and there are several other definitions of emittance in the literature, some referring to the area of an ellipse in phase space, others without the factor of $\sqrt{U}$ (i.e. unnormalized), and some in a relativistic form, most useful for electrons in high-energy accelerators.[65,66]  Also, it is important to recognize that the conservation of normalized emittance only holds in the absence of Coulomb interactions between the ions and when aberrations in the focusing optics can be ignored. While these phenomena often cause very significant effects, it is useful to neglect them, at least at the outset, in establishing a framework for describing the beam.

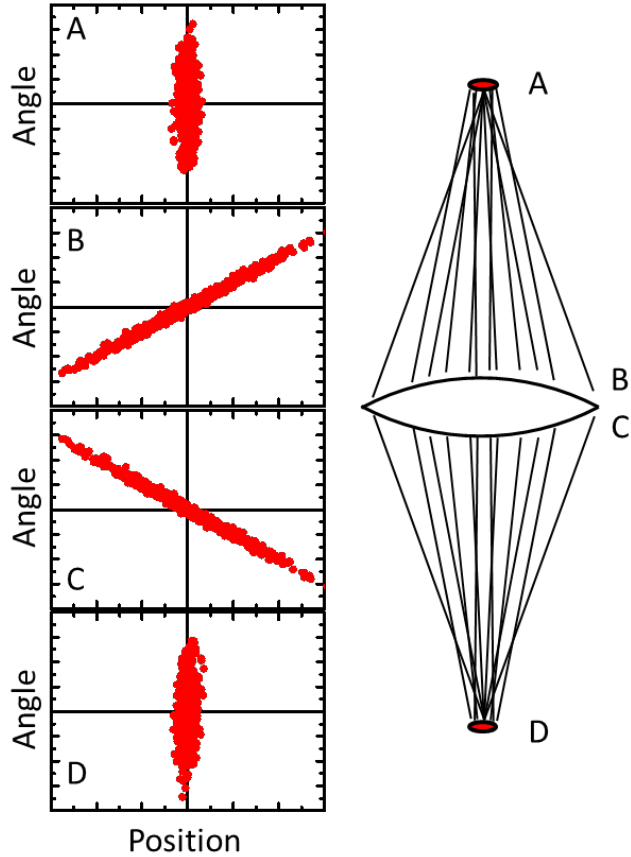

FIG. 1. Evolution of emittance as a beam is focused. Each dot represents the position and angle coordinate of a trajectory. At the source (A), position and angle are not correlated. After propagating to just before the lens (B), correlation between position and angle develops. After passing through the lens, the angles are reversed (C). On propagating to the image position (D), the correlation between position and angle disappears.

At the location in a source where ions are first created, it is often a good assumption that there is no correlation between position and angle in the beam and the emittance simplifies to

$$\varepsilon_x = \sigma_x \sigma_\alpha \sqrt{U}, \tag{3}$$

where $\sigma_x$ is the RMS deviation of trajectory positions, $\sigma_\alpha$ is the RMS deviation of trajectory angles, and $U$ is the average beam energy along the axis. For a source where the temperature is the dominant determinant of transverse motion, this further reduces to

$$\varepsilon_x = \sigma_x \sqrt{k_B T/2}\,, \tag{4}$$

where $k_B$ is Boltzmann's constant and $T$ is the source temperature.

Figure 1 gives a visual representation of the evolution of the emittance as the beam propagates. At the source location, where there is no correlation between position and angle, the distribution of ion trajectories can be represented by an elliptical cloud in position-angle space with major and minor axes oriented along the $x$ and $\alpha$ directions. As the ion beam propagates away from the source region, the ellipse narrows and rotates as the correlation term in the emittance grows. If the ions pass through a lens, the ellipse is flipped about the $x$ axis as the trajectories are redirected with new angles. After some propagation distance, an image of the source is formed, and the elliptical distribution of trajectories again becomes rectilinear. At this point the correlation between position and angle disappears, and the emittance again simplifies to Eq. (3).

*Contributions to spot size*

Because Eq. (3) is valid anywhere an image of the source is formed, in particular at the focus of the final lens in a FIB system, it provides a simple way to estimate the contribution to the focal spot size that arises from inherent properties of the source. As long as emittance is conserved, we can write

$$\sigma_x' = \frac{\varepsilon_x}{\sigma_\alpha' \sqrt{U}}\,, \tag{5}$$

where $\sigma_x'$ and $\sigma_\alpha'$ are the RMS position and angular spreads at the focus. Equation (5) shows that the emittance of the source has a direct impact on the attainable source size at the focus of a system. It also indicates that a smaller spot can generally be obtained by increasing $\sigma_\alpha'$ and increasing $U$, assuming no other influences (such as aberrations) are limiting the resolution.

While Eq. (5) provides a simple expression for the spot size in a FIB, it is important to realize that emittance is only one contributor to the ultimate resolution. In fact, assuming the beam energy is set by other requirements for the system, it is natural to try to make $\sigma_\alpha'$ as large as possible, and to ask what the limitations are in increasing this angular spread. Aside from some practical constraints based on attainable beam sizes and working distances, a critical factor limiting $\sigma_\alpha'$ is the fact that all ion optical lenses have aberrations, and these generally get larger with increasing $\sigma_\alpha'$. Because aberrations increase and the emittance contribution decreases with larger $\sigma_\alpha'$, an optimum condition exists when the contributions from each are roughly equal. As a result, the performance of a FIB system cannot be fully characterized without some attention to aberrations.

The primary aberrations that play a role when focusing a beam to a small spot are chromatic and spherical. Chromatic aberration results when ions with different longitudinal energies focus differently in a given lens, and spherical aberration arises when the focal length depends on the ion trajectory's distance from the axis. While a great deal of detail has been worked out for treating aberrations in charged particle optics (see e.g., ref. 67), for present purposes we can concentrate on the behavior of a nearly parallel beam being focused to a small spot. In this case the chromatic and spherical aberration contributions to the focal spot size can be written as

$$d_c = \xi \, C_c \, \Delta\alpha \, (\Delta U/U) \qquad (6)$$

$$d_s = \eta \, C_s \Delta\alpha^3, \qquad (7)$$

where $d_c$ and $d_s$ are measures of the beam diameter, $\xi$ and $\eta$ are numerical constants, $C_c$ and $C_s$ are the chromatic and spherical aberration coefficients, $\Delta\alpha$ is a measure of the convergence angle spread, and $\Delta U/U$ is a measure of the relative energy spread in the beam. The values of the numerical constants $\xi$ and $\eta$ depend on the definitions of $d_c, d_s, \Delta\alpha$, and $\Delta U/U$, and also on the form of the distributions in $\alpha$ and $U$. The aberration coefficients are properties of a given lens geometry, and are usually derived from detailed numerical ray tracing calculations, or occasionally by experimental determinations.

A common measure for the beam diameter is $d_{50}$, the diameter that contains 50 % of the beam current. Taking this definition for $d_c$ and $d_s$, and assuming a uniform circular distribution in $\alpha$ up to a fixed angle $\alpha_0$ and a Gaussian distribution in $U$ with full width at half maximum (FWHM) $\Delta U_{1/2}$, it can be shown that[68,69]

$$d_c = 0.34 \, C_c \, \alpha_0 \, \frac{\Delta U_{1/2}}{U} \qquad (8)$$

$$d_s = 2^{-5/2} \, C_s \alpha_0{}^3. \qquad (9)$$

These spot size contributions can be compared with the spot size arising from beam emittance if we assume the emittance contribution has a Gaussian distribution at the focus. In this case the 50 % beam current diameter $d_\varepsilon = 2\sqrt{2\ln 2} \, \sigma'_x$ and $\sigma'_\alpha = \alpha_0/\sqrt{2}$, resulting in

$$d_\varepsilon = \frac{4\sqrt{\ln 2} \, \varepsilon_x}{\alpha_0 \sqrt{U}}. \qquad (10)$$

The expressions in Eqs. (8-10) are presented here as examples because they are in common usage in the literature, and correspond to angular and energy distributions found in systems based on point sources like the LMIS and GFIS. It should be noted that for the cold atom sources discussed in this review, distributions will generally be different, resulting in different values for the coefficients $\xi$ and $\eta$.

When all contributions to the spot size are roughly equivalent, i.e., where the net spot size is near a minimum, it is useful to consider how the various contributions combine. Generally speaking, the exact form of the combination depends on the distributions in $\alpha$ and $U$, though one might expect some form of root-power-sum would be appropriate in order to preserve dominance by any one contribution when it becomes the largest, and also to ensure smooth variation between contributions. For the case of a uniform distribution in $\alpha$ and a Gaussian distribution in $U$, Barth and Kruit[68] have shown that the net spot size $d_{50}$ is well represented by

$$d_{50} = \left[ (d_\varepsilon^{1.3} + d_s^{1.3})^{2/1.3} + d_c^2 \right]^{1/2} \tag{11}$$

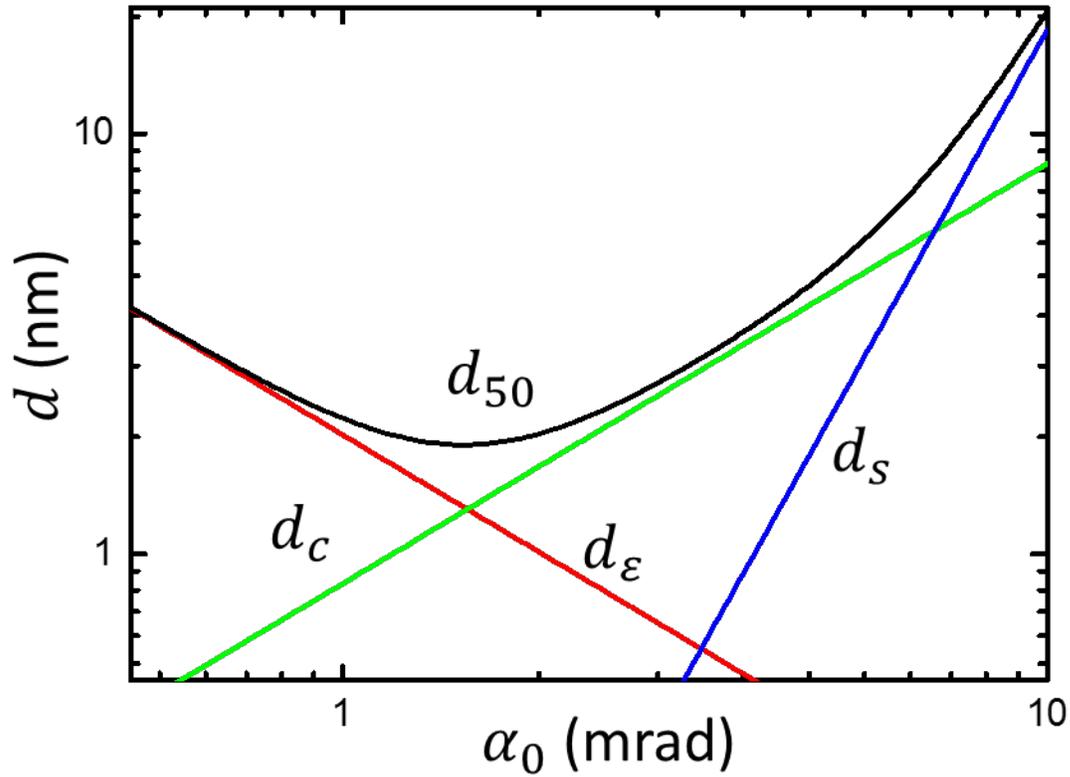

FIG. 2. Combination of aberrations as a function of convergence angle at the focus, $\alpha_0$. The three contributions to source size, $d_\varepsilon$ (emittance), $d_c$ (chromatic aberration) and $d_s$ (spherical aberration) combine according to Eq. (11) to yield a net spot diameter $d_{50}$, defnied as the diameter containing 50 % of the beam current. Shown is a calculation for a hypothetical ion beam with $\varepsilon = 10^{-10}$ m rad $\sqrt{\mathrm{eV}}$, $U = 3 \times 10^4$ eV, $\Delta U_{1/2}/U = 10^{-4}$, $C_c = 25$ mm, and $C_s = 100$ mm.

Using Eq. (11), it is possible to find the minimum spot size that can be expected given a source with a specified emittance $\varepsilon_x$, a beam energy $U$ at the focus, a convergence angle $\alpha_0$, an energy spread $\Delta U_{1/2}$, and aberration coefficients $C_c$ and $C_s$. In Fig. 2 we illustrate how the various components in $d_{50}$ contribute as a function of $\alpha_0$ with a hypothetical example of an ion beam with $\varepsilon = 10^{-10}$ m rad $\sqrt{\mathrm{eV}}$, $U = 3 \times 10^4$ eV, $\Delta U_{1/2}/U = 10^{-4}$, $C_c = 25$ mm, and $C_s = 100$ mm. While Fig. 2 is useful for understanding how a minimum spot size is always reached when both aberrations and emittance make significant contributions to the spot size, it is important to recognize that $C_c$ and $C_s$ are both generally strong functions of the focal length of a lens, and so if $\alpha_0$ is varied by changing the focusing power of the objective lens the dependence is not as simple as shown in Fig. 2.

*Brightness*

While emittance provides a useful characterization of a source in terms of its potential for forming a high resolution focused beam, it does not alone provide enough information to address the problem of getting as much current as possible into a given focal spot size, or getting a useful amount of current into the smallest spot possible. In fact, the emittance of a beam can be reduced almost at will by introducing apertures in the ion optical system to remove ions with high-emittance trajectories. The result of this is not only a reduction in emittance, but also a reduction in the beam current, which while necessary in some circumstances, is generally not desirable and defeats the main goal of getting as much current as possible into the spot.

A characterization of the source that takes into account the current that can be supplied is the normalized (or reduced) brightness, which is defined as the current density per solid angle per beam energy.[65] Formally, the normalized brightness is defined as

$$B = \frac{1}{U} \frac{\partial^2 I}{\partial A \partial \Omega}, \tag{12}$$

where $I$ is the beam current, $A$ is the cross-sectional area of the ion beam, $\Omega$ is the solid angle, and $U$ is the beam energy. For a thermal source at temperature $T$, the angular distribution is uniform into a solid angle of $2\pi$ in the forward direction, and the average beam energy along the $z$-direction is $kT/2$. In this case the normalized brightness reduces to a relatively simple form:

$$B(x,y) = \frac{J(x,y)}{\pi k_B T}, \tag{13}$$

where $J(x,y)$ is the current density in the plane of emission. In Eq. (13) the normalized brightness and current density are written with explicit dependence on $x$ and $y$ because the current density can vary across a source. Often, however, when it is desirable to discuss "the" brightness of a source, it is useful to use the peak normalized brightness $B_{peak} = B(0,0)$ (assuming a symmetric distribution peaked at the center), or a spatially averaged normalized brightness.

Using the fact that the emittance contains information about the spatial and angular spreads as well as the energy of the beam, it is possible to write the normalized brightness in terms of the emittance. The exact expression depends on the form of the distributions in space and angle, but if we assume a Gaussian spatial distribution and a uniform angular distribution, we can use Eq. (4) and Eq. (13) to write the peak normalized brightness of a source as

$$B_{peak} = \frac{I_0}{4 \pi^2 \varepsilon_x \varepsilon_y}, \tag{14}$$

where $I_0$ is the total current in the beam, and $\varepsilon_x$ and $\varepsilon_y$ are the normalized emittances along the two axes perpendicular to the beam axis (note, by comparison, the normalized brightness for a uniform circular distribution in space is a factor of two smaller).

The normalized brightness is a very useful figure of merit for a beam because it contains all the information in the emittance, plus a measure of the current in the beam.  Furthermore, it has the valuable property that in many situations (e.g., if the current density is uniform) it is not changed by placing apertures in the beam, since the numerator and denominator of Eq. (14) both scale as the square of the radius of any aperture introduced.  This property makes normalized brightness one of the most important ways to characterize a source, since it is often decoupled from the specific ion optical system used to make a focused ion beam.

Assuming the normalized brightness doesn't change in a beam, Eq. (13) can be used together with Eq. (14) to derive an expression for the beam current expected at the focus:

$$I_0 = 4\,\pi^2 \varepsilon_x \varepsilon_y B = 4\,\pi \frac{\varepsilon_x \varepsilon_y J_0}{k_B T}, \tag{15}$$

where $\varepsilon_x$ and $\varepsilon_y$ are the final emittances in the beam, after any apertures or other perturbations, and $J_0$ is the peak current density in the (assumed Gaussian) source.

With eqs (11) and (15), knowledge of the emittance, normalized brightness, and/or total current of the source, and estimations of the aberrations expected in the ion optics to be used, it is possible to estimate the spot size and current that could potentially be obtained in a focused ion beam system.  However, it should be recognized that there are other phenomena that can play a large role, and actual performance can be significantly less than expected in some circumstances.

The most significant phenomenon that can impact performance is Coulomb interactions between the ions in the beam.  Coulomb interactions affect the properties of ion beams in two ways, via space charge, and through stochastic collisional effects.  Space charge arises when there is sufficient charge density in the beam for ions travelling down the beam to be presented with a potential "hill" formed by the aggregate charge in front of them.  This has the effect of defocusing the beam and results in aberrations and a reduction of the normalized brightness.  Stochastic effects are the consequence of many random interactions between the ions. These interactions can be very significant even in low density beams because of the long range nature of the Coulomb interaction.  The result can be a substantial heating of the beam, and transfer of temperature from longitudinal to transverse degrees of freedom.  This also has a deleterious effect on the emittance and normalized brightness of the beam.

Much theoretical work has been done to analyze the effects of Coulomb interactions in a beam.[65,70,71,72,73,74] Analytical models have tended to be quite complex because of the long range nature of the interactions and the interplay between beam density and the magnitude of the effect.  As an alternative, a numerical approach can be taken, using Monte Carlo simulations and calculating the interactions between all ions in the beam individually for each ion, or discretizing the charge distribution.  Several successful approaches in this manner have been described,[70,75,76,77,78,79,80] and some of the methods are now available in commercial software packages.

The understanding of inherent source capabilities provided by emittance, brightness and energy spread, together with knowledge of the ion optical system with its aberrations, plus a sense of how important

Coulomb interactions will be, provide a good basis for estimating how well a focused ion beam system will perform. Once it has been established that a source has potential for high performance, the next step is to design an optimized system. For this, the best approach is to apply detailed numerical ion optical calculations, which, with today's level of sophistication, can provide quite accurate predictions, allowing refinement of the design to the point of obtaining the best possible performance.

## IV. Conventional FIB Sources

The three most common ion sources used in focused ion beam applications today are the liquid metal ion source (LMIS), the gas field ionization source (GFIS), and the inductively coupled plasma source (ICP). The basis of operation of these sources has been extensively covered in the literature and will only be summarized here. Our main purpose is to show how the challenge of getting as many ions as possible into as small a spot as possible has been approached in the past, in order to provide some context for the cold atom ions sources that are the subject of this review.

*Liquid metal ion source*

Of the three conventional ion sources, the LMIS is by far the most widespread, due in large part to its simple construction, its stable, repeatable performance, and its high brightness. A number of commercial systems are available, and these have found their way into research labs, nanocenters, and industrial production lines around the world. The LMIS consists of a sharp tip connected to a small reservoir of metal.[1,26,30,81] The metal is generally chosen to have a low melting point and a very low vapor pressure, allowing it to remain molten during operation, yet not evaporate excessively. This combination is relatively rare, but gallium is a good example of a metal with these properties, with a melting point just above room temperature, and a vapor pressure of less than $10^{-9}$ Pa at this temperature. Some alloys are also suitable, when used with modest heating of the source, expanding the number of ionic species that can be created.[6] The reservoir is arranged so that the liquid metal can flow to the tip and wet the sharp apex. A relatively high voltage of a few thousand volts is applied to the tip, creating a strong electric field that draws the liquid metal into a very sharp cone, known as a Taylor cone.[82] At the tip of the Taylor cone the electric field is strong enough to cause field evaporation of metal ions, which are then accelerated and formed into a beam.

For a $Ga^+$ LMIS, the emission current is typically in the range of 2 μA to 10 μA, and the angular spread is in the range 25° to 35° FWHM. The source size is a somewhat complex quantity to pin down because the shape of the Taylor cone depends on the current and extraction voltage, and Coulomb interactions have a very strong influence at the point of emission. Nevertheless, an effective source size can be determined, and this has been studied extensively.[83] Consensus is that the ions come from an effective source of about 50 nm diameter. The normalized brightness of the $Ga^+$ LMIS has been measured[84] to be as high as $1 \times 10^6$ A m$^{-2}$sr$^{-1}$eV$^{-1}$, and the energy spread in the $Ga^+$ LMIS is typically in the range of 4 eV to 5 eV. This relatively large energy spread is a result of the field evaporation process in combination with very strong Coulomb interactions at the emission point.

While the $Ga^+$ LMIS emits microamperes of current, a high resolution beam is only obtained when an aperture is used to select a small fraction of the beam at the center, reducing the emittance to a level commensurate with the desired spot size. In addition, the ions are usually accelerated to a few tens of kiloelectronvolts to minimize the effects of chromatic aberration and take advantage of the inverse root energy dependence in the spot size (see Eq. 5). In high resolution mode, spot sizes in the range 5 nm to 10 nm can be obtained with currents of 1 pA to 2 pA at a beam energy of 30 keV.

Although the $Ga^+$ LMIS has proven extraordinarily useful for many applications, the application space for FIBs is so broad that there are a number of situations where it is not ideal. For example, it is sometimes desirable to have an ion species other than gallium. To address this, it is possible to use other metals or alloys in an LMIS. It has generally proven difficult, however, to obtain the same level of stability and repeatability seen with gallium with other species, and the desired species may not be among those for which a source is possible. In addition, these sources require a mass filter to select the desired species when an alloy is used, which not only increases the complexity of the instrument, but also can cause increased aberrations in the beam, and often results in reduced brightness because it is necessary to discard the undesired ions. In other applications the requirement may be for a low energy ion beam. In this case, the LMIS energy spread can be problematic, because it is fixed at a relatively high value. When the beam energy is lowered, the ratio of the spread to the mean beam energy is reduced, causing increased chromatic aberration (see Eq. 6). Together with the increased emittance contribution to the spot size from the lower energy, this can result in an unacceptable focal spot size.

*Gas field ionization source*

Though the principle of the gas field ionization source has been known since the 1970s,[2] it has not been until recently[85] that it has emerged as a reliable option for FIBs with some exceptional properties. Like the LMIS, the GFIS relies on the very strong electric field that arises when a high voltage is applied to a sharp tip. In the GFIS, however, the tip is specially treated to create a trimer of single atoms at the apex, resulting in a very high field concentrated on the surface of these atoms. When a small amount of gas is introduced to the tip, ionization of the gas atoms occurs in the high electric field, and with proper choice of voltage, this can be restricted to happen only on the surface of the trimer atoms. The result is ion emission essentially from a single atom, which translates into an extremely high brightness.

The GFIS has seen most of its implementation using He gas. Because He is a noble gas and has a very high ionization potential, it has proven to be the easiest to use for creating a stable source. Having a very low mass, it is also very desirable for ion microscopy applications, where sputtering is to be minimized. Using He, the GFIS has been measured[86] to have a normalized brightness of $10^9$ Am$^{-2}$sr$^{-1}$ eV$^{-1}$, an energy spread of 1 eV, and a probe size of 0.35 nm at an energy of 30 keV. Typical currents at the probe are 0.1 pA to 10 pA. Recently, a Ne ion beam has also been demonstrated using a GFIS,[4,87,88] with similar brightness and energy spread.

With its extraordinarily high brightness and its corresponding small probe size, the GFIS FIB has proven very useful for a number of applications. Using He ions for microscopy, very high quality images have been obtained from a wide range of samples. Also, a number of recent studies have shown that $He^+$ is well suited to milling applications where very small apertures in thin membranes are desired.[89,90,91]

While the GFIS has proven to have excellent performance from the point of view of getting as much current as possible into as small a focal spot as possible, it nevertheless has some drawbacks for some applications. For example, because of the high field necessary to extract the ions, it is most practical to implement with a relatively high beam energy at the focus. This makes it difficult to create a low energy focused beam without extraordinary measures such as floating the sample stage at a high positive potential. Also, the number of ionic species accessible is quite limited, most likely just to the lightest noble gases, since a high ionization potential is required and the tip is very sensitive to chemical attack. In addition, there is no possibility of a high-current, low-resolution mode, which could be useful, for example, in large scale milling applications.

*Inductively coupled plasma source*

Unlike the LMIS and the GFIS, the inductively coupled plasma source does not obtain its ions from a sharp tip, but rather from a more conventional approach of ionizing a gas via electron currents induced in a plasma.[92] Typically a plasma is created in a chamber with a gas inlet and a small (0.1 mm to 0.2 mm) exit orifice by subjecting it to a strong radio-frequency (RF) field created by an antenna wrapped around the exterior. The RF frequency is chosen to be high enough to be well above the plasma ion resonance, but also below the electron resonance. Ions are created by electron impact and extracted through the exit orifice at a few kiloelectronvolts.

The brightness of the ICP source does not derive so much from a small source size, as it does for the LMIS and GFIS, as from a large current and relatively small angular spread. Typical values for the normalized brightness range from $4 \times 10^2$ $Am^{-2}sr^{-1}eV^{-1}$ for $O^-$ ions to $1 \times 10^4$ $Am^{-2}sr^{-1}eV^{-1}$ for $Xe^+$. While these brightness levels do not match the LMIS or GFIS, they are still sufficient for several important applications. With a typical energy spread of 3.5 eV to 5 eV, ICP-based focused beams have been demonstrated with 50 nm to 200 nm spot size at 15 keV and 1 pA for $O^-$, and 60 nm spot size at 30 keV and 0.1 nA for $Xe^+$.[93] The $O^-$ FIB is of particular importance for SIMS applications where electropositive elements are to be analyzed,[94] and the high current $Xe^+$ beam is particularly attractive for rapid milling of large volumes with low contamination.

The strongest advantages of the ICP source are its flexibility in producing different ionic species, and its ability to produce high currents at reasonable spot sizes. Due to its lower brightness, however, it is not suitable for the most demanding high resolution nanoscale applications.

## V. Cold-atom-based sources

Considering the three types of conventional FIB sources discussed in section IV, it is evident that the approach to high brightness has historically taken either of two paths: reduce the emittance of the source by reducing the spatial spread $\sigma_x$ in Eq. (4), or simply increase the current in order to increase the numerator in the normalized brightness, Eq. (14). By examining Eq. (4), however, it is clear that there is another way to increase the normalized brightness, i.e., by reducing the temperature of the source. This approach has not been particularly fruitful for conventional sources, because it is generally difficult to make a significant change in the source temperature by more than two orders of magnitude, for example, by cooling with liquid helium from 300 K to 4 K. Since the emittance only scales with the square root of the temperature, this only results in a factor of ten improvement in the emittance. This limited improvement, together with the significant additional complexity associated with source cooling, makes this approach somewhat unattractive (although it is worth noting that some earlier GFIS systems did have cryogenically cooled tips, and modern ones use liquid nitrogen cooling). It is this approach, however, that is taken by the cold atom ion sources discussed in this review. With the extraordinarily low temperatures attainable through laser cooling, the option of reducing the emittance by lowering the temperature has become a realistic choice.

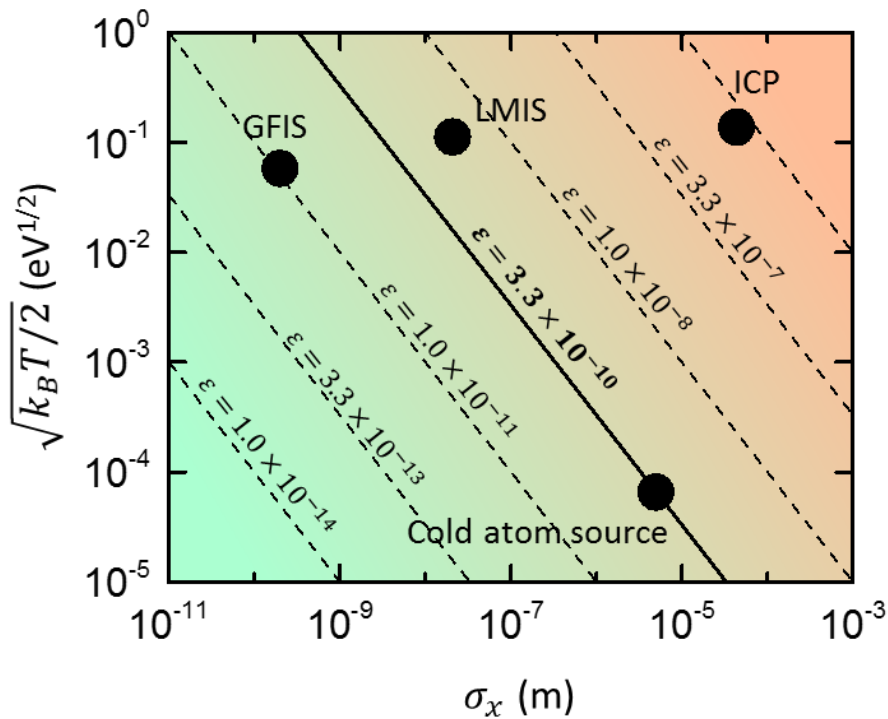

FIG. 3. Source emittance for the gas field ionization source (GFIS), liquid metal ion source (LMIS), inductively couple plasma (ICP) source, and cold atom source. Each source is characterized by a point on the graph according to its size $\sigma_x$ and temperature $\sqrt{k_B T/2}$. The emittance $\varepsilon$ is given by Eq. 3. Lines of constant emittance go diagonally from upper left to lower right.

As will be discussed in detail in section VI below, laser cooling can easily produce temperatures as low as a few hundred microkelvin, and often can result in much lower values. Thus the effective source temperature can be reduced from room temperature by over six orders of magnitude. This reduction has a dramatic effect on the emittance of a source. For example, consider a temperature of 100 μK and an ionization region diameter of, say, 10 μm. Eq. (4) then results in an emittance of $\varepsilon = 3.3 \times 10^{-10}$ m rad $\sqrt{eV}$. This value is similar to, or even somewhat smaller than, the emittance of a Ga⁺ LMIS operated in high resolution mode.[95] This suggests that, at least from an emittance point of view, a cold-atom-based ion source should be able to provide a focal spot in the nanometer range at reasonable beam energies without significantly reducing the beam current with an aperture.

To illustrate the different ways that source size and temperature contribute to emittance, Fig. 3 shows a plot of $\sqrt{k_B T/2}$ vs. $\sigma_x$ for LMIS, GFIS, ICP and cold atom sources. In this plot, lines of constant emittance lie diagonally from upper left to lower right. As seen in the figure, the GFIS, LMIS and ICP sources have different emittances, but the variation is governed almost entirely by the source size with very little variation in temperature. The cold atom source, on the other hand, attains a low emittance by drastically lowering the temperature instead.

Once a population of cold atoms is created, to make an ion source it is necessary to ionize them without heating by any significant amount. A convenient way to do this is through photoionization. As long as the ionization laser energy is very close to the ionization threshold energy, the ion receives very little excess energy in this process and the ions do not heat up appreciably. This is helped by the mass ratio of the electron to the nucleus, which ensures, via conservation of energy and momentum, that most of the kinetic energy is carried away by the electron.

If the photoionization is done in the presence of an electric field (as is usually the case), a potential gradient is imposed on the atom, and the ionization threshold becomes spatially dependent. Electrons escape more readily on the low energy side of the atom, and are less likely to exit on the high energy side. The result is a rather complex ionization spectrum that does not have a well-defined threshold, but instead transitions gradually from a series of autoionizing Rydberg states to a continuum as the photon energy is varied from below the "saddle point" at the front of the atom to above the field-free ionization limit. Within this range, the kinetic energy of the electron that leaves the atom can be a complex function of the photon energy because of the variety of quantum states that exist in this region.[96] This can be an issue for cold atom-based electron sources,[13] but for ions it is less likely to be problematic because the mass ratio of the nucleus to the electron usually ensures the ions will not be significantly heated.

Typically, it is most practical to use a two-step ionization process, in which the atom is first excited to an intermediate state, and then further excited to the ionization threshold. A practical reason for this is that ionization thresholds are usually several electron volts above the ground state, and would require a deep UV laser to be accessed directly. It is also useful to use a two-step process because it can help limit the ionization to only atoms that have been cooled.

The current that can be produced by a cold-atom ion source is generally dependent on how the source is implemented, and more detail will be given in sections VII and VIII, where specific sources are discussed. For the present, it is instructive to discuss a rough estimate of what the current could be by considering the flux of atoms that can generally be produced by an atomic beam source, as this will impose an absolute maximum on the current. Typically, an atomic beam can produce atoms at a rate of $10^9$ s$^{-1}$ to $10^{10}$ s$^{-1}$. Assuming 100 % ionization efficiency, this corresponds to a current of 0.2 nA to 2 nA. Combining this current with the emittance estimate above, Eq. (14) can be used to generate an upper limit normalized brightness estimate of well over $10^7$ A m$^{-2}$sr$^{-1}$eV$^{-1}$. This is significantly higher than the normalized brightness of the $Ga^+$ LMIS source, suggesting that cold atom ion sources could in principle be very attractive. We note, though, that other limitations on the current do exist in some situations, and these will be discussed below.

In addition to having a brightness that potentially rivals the brightest sources available today, cold-atom ion sources have a number of potential advantages stemming from the unique way in which their brightness is generated. Because the brightness arises from a low temperature, instead of a small source size, cold atom ion sources tend to have a relatively large source area and a very narrow angular spread. Ion optically, this corresponds to a virtual source located a long distance behind the physical source position. Such a configuration results in a very high demagnification factor when the ion beam is focused, in some cases as high as 2500x. The result of this is a much-reduced sensitivity to vibrations or spatial instabilities associated with the source.

Another result of the large spatial extent of the source is a lack of any region of high current density. Whereas in a tip-based source, all the current must pass through a nanometer-scale region as the ions are emitted, the cold atom ion source produces current over a much broader area. The resulting current density can be a factor of order $10^4$ lower, and this can lead to a reduced role played by Coulomb interactions. Since Coulomb interactions depend on the current density, a cold atom ion source's brightness could in principle be less affected by those interactions. Beam energy also influences Coulomb interactions, however, and since the electric field strength in a cold atom source is generally less than in a tip source, Coulomb heating cannot be neglected entirely. We discuss this at greater length in section VIII below.

A consequence of reduced Coulomb interactions in cold atom ion sources is an energy spread that can be much smaller than what is typically found in other ion sources. Most sources, with a high current density at the point of ion creation, are subject to anomalous energy broadening (also known as the Boersch effect),[97] which originates from longitudinal heating of the beam due to Coulomb interactions. This effect, in combination with an inherent emission-related energy spread, can lead to energy spreads as high as 5 eV for the $Ga^+$ LMIS source. By contrast, the energy spread in a cold atom ion source can be quite small. The fundamental limit, based on the source temperature, is extremely small, of order a few nanoelectronvolts. However, a practical limit is imposed by the need to ionize in the presence of an electric field, which is necessarily associated with a potential gradient. Whatever extent the source has along the potential gradient will result in a range of energies for the ions produced. While this spread can be engineered to be large or small, depending on the needs of a particular application – and can in fact be compensated by pulsing the ionization and ramping the field[98] – it generally works out to be in the range of a few tenths of an electron volt. This is much larger than any anomalous broadening that might occur, and so is the dominant contribution to the energy spread. It is significantly smaller than the spread found in most other ion sources, giving the cold atom ion source a unique insensitivity to chromatic aberrations in the ion optical system used to accelerate the ions and create the focused ion beam. This makes design of the ion optical system easier, and also opens new possibilities for producing highly focused low energy ion beams.

Another advantage that cold atom ion sources provide is potential access to a whole new group of ionic species. As will be discussed in section VI, laser cooling has been demonstrated for over 27 atoms in the periodic table, and these are species typically not easily produced in other high brightness ion sources. While each atom requires its own specific cooling and ionization laser wavelengths, the principles of source operation are generally the same for all species. It should be noted, however, that achievable current densities and temperatures depend on specific atomic transition strengths and the availability of lasers with sufficient power and linewidth control. These can vary significantly between atomic species, and thus high brightness operation may not be possible for all laser-coolable atoms.

An interesting consequence of the spectroscopic nature of the laser cooling process is that cooling in one species generally has no effect on other species. This suggests the possibility of combined sources with control over the relative composition of the beam. This control extends to individual isotopes, since the laser cooling process is sensitive to isotope shifts in the atomic transition frequency. As a result, any cold atom ion source will be inherently isotopically pure, and different isotopes can be selected by a simple change in the laser frequency tuning.

A final advantage of cold atom ion sources worth mentioning is the degree of control afforded by the laser-based photoionization process. Because the photoionization laser power can be varied while keeping all other parameters the same, the beam current can be varied without any change in extraction field or geometry. This enables a great deal of flexibility in optimizing the brightness and energy spread for specific applications – a flexibility that is generally not possible with LMIS, GFIS or ICP sources. Moreover, using simple optical methods, it is relatively straightforward to create complex intensity distributions in the ionization region, allowing unique control over the spatial distribution of ion creation. This has been used, for example, to create patterned beams in both a cold-atom ion source[99] and a cold-atom electron source.[12] Also, the great timing flexibility of laser systems can be brought to bear, for example with laser pulses as short as femtoseconds, or complex temporal control of intensity with nanosecond resolution through acousto-optical or electro-optical modulators. This flexibility opens many opportunities for timing experiments, such as lock-in detection of small signals or time-of-flight analysis of scattered ions. These would generally be difficult with conventional ion sources, because the opportunities for modulating the ion beam are limited.

## VI.    Laser cooling and trapping

Before discussing specific implementations of cold atom ion sources, we provide a brief review of laser cooling and trapping to provide some understanding of how these techniques enable such extremely low temperatures. Since the beginnings in the late 1970s and early 1980s, laser cooling and associated trapping techniques have led to many breakthroughs in atomic, molecular, and optical (AMO) physics, and a number of reviews on the subject have been published.[100,101,102,103,104,105,106] Although probably best known for their use in modern atomic clocks to enable the highest possible accuracy in time and frequency standards,[107,108] laser cooling and trapping techniques have also been used to create new quantum states of matter such as Bose-Einstein condensates[109] and Fermi degenerate gases.[110] Outside of AMO physics, laser cooling and trapping has proven to be useful as a platform to perform experiments in a broad range of scientific areas. Laser cooled atoms have been used for experimental simulation of complex condensed matter systems,[111] nanofabrication,[112] creation of ultracold plasmas,[113] as a possible platform for quantum computing,[114] and for creating high-brightness FIBs as discussed in this review.

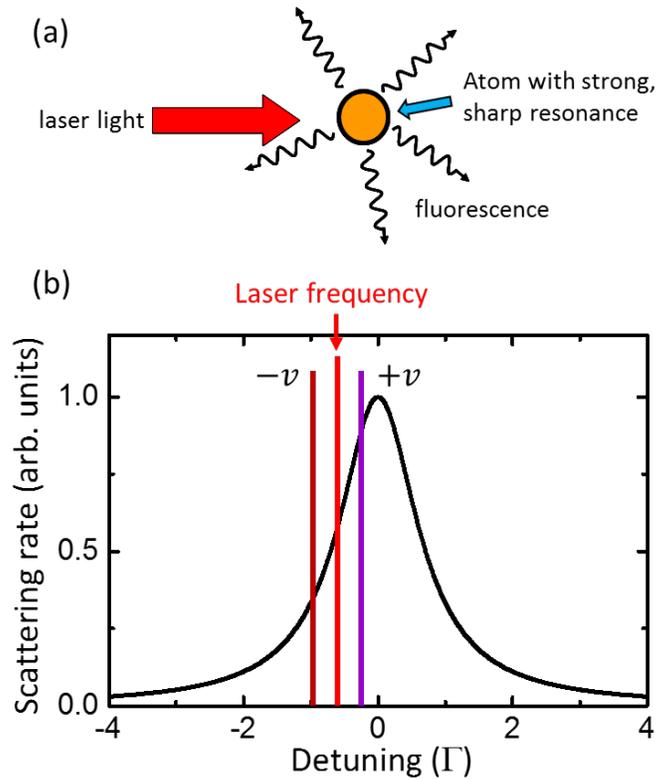

FIG. 4. (a) Light force. Laser light with momentum in a single direction is absorbed by an atom and re-radiated in all directions, resulting in a net transfer of momentum to the atom. (b) Laser cooling. The scattering rate is shown for an atom at rest with a laser tuned to the red of atomic resonance. For atoms with a velocity ($+v$) in a direction opposite to the laser direction, the Doppler shift increases the apparent laser frequency and the laser light is absorbed more strongly compared to atoms travelling with velocity $-v$. The result is a velocity-dependent force that damps the atomic motion.

*Laser cooling*

Laser cooling uses the momentum-carrying properties of light to change the kinetic energy of target atoms. When an atom absorbs a photon of wavelength $\lambda$ from a laser, the momentum of the incoming photon, $h/\lambda$ ($h$ is Planck's constant), is transferred to the atom, giving it a momentum kick in the direction of laser propagation. When the atom subsequently spontaneously emits a photon as it decays back to the ground state, it receives another momentum kick as it recoils, but this time in a random direction. On undergoing a number of directed-momentum absorptions followed by random-direction emissions which average to zero, the atom experiences a cumulative average net momentum transfer per scattered photon equal to $h/\lambda$. The result is the so-called "light force" on atoms (see Fig. 4a).

While the light force can in general either add to or subtract from the target atom's kinetic energy, laser cooling can only take place when a situation is created where kinetic energy is removed in a dissipative way, resulting in a cooling effect. This is accomplished by taking advantage of two characteristics of light scattering: the strong dependence of the scattering rate on the detuning of the laser light from the atomic resonance, and the Doppler effect, which creates a velocity-dependent detuning. For a two-level atomic system illuminated by laser light with intensity $I$, detuned by a frequency $\Delta$ from the atomic resonance, the scattering rate is given by[106]

$$\gamma_p = \Gamma \frac{I/2I_{sat}}{1 + I/I_{sat} + 4\Delta^2/\Gamma^2},$$  (16)

where $\Gamma$ is the spontaneous decay rate of the transition and $I_{sat} = \pi hc\Gamma/(3\lambda^3)$ is the saturation intensity, an inherent property of the specific atomic transition being utilized ($c$ is the speed of light). Fig. 4b shows a plot of the scattering rate, illustrating how it decreases sharply with detuning either above or below the atomic resonance. The Doppler shift for an atom traveling with velocity $v$ is given by $\pm 2\pi v/\lambda$, where the positive sign is for atoms traveling in a direction opposite to the laser propagation, and the negative sign is for atoms traveling along the laser propagation direction.

Laser cooling occurs when the laser is tuned to the red of the atomic transition (lower frequency), on the side of the Lorentzian profile. In this case an atom moving opposite to the direction of laser propagation will experience a Doppler shift that increases the apparent frequency of the laser, exciting the atom closer to the atomic resonance. The atom will then feel a correspondingly stronger light force opposing the motion. The result is a deceleration of the atomic motion by a force that increases with increasing speed. Such a velocity-dependent force creates the condition for dissipation, or damping of the motion, and the consequence is a removal of kinetic energy, which is carried off by the photons scattered into random directions. If red-detuned lasers are counter-propagated from all directions, atoms in the laser field will always have a velocity-dependent light force that opposes their motion regardless of their direction of travel. The resulting damping of the atomic velocity is known as Doppler cooling, and the region of space where the laser beams overlap is often referred to as "optical molasses".[115]

The Doppler cooling force for a two-level atom in one-dimension as a function of velocity is given by[106]

$$F_{Dop} = \frac{h\Gamma}{2\lambda}\left(\frac{I/I_{sat}}{1 + I/I_{sat} + 4\left(\frac{\Delta - 2\pi v/\lambda}{\Gamma}\right)^2} - \frac{I/I_{sat}}{1 + I/I_{sat} + 4\left(\frac{\Delta + 2\pi v/\lambda}{\Gamma}\right)^2}\right).$$  (176)

A plot of Eq. (176) is shown in Fig. 5. This figure shows that for sufficiently small velocities, atoms feel a damping force roughly proportional to their velocity and thus experience cooling. It also highlights a significant characteristic of laser cooling: there is a "capture range" of velocities within which cooling can take place, and beyond which cooling becomes ineffective. This range is roughly given by the location of the peaks in Fig. 5, which occur at $v_c = \pm\lambda\Delta/(2\pi)$. For most atoms, $v_c$ is of order a few tens of meters per second.

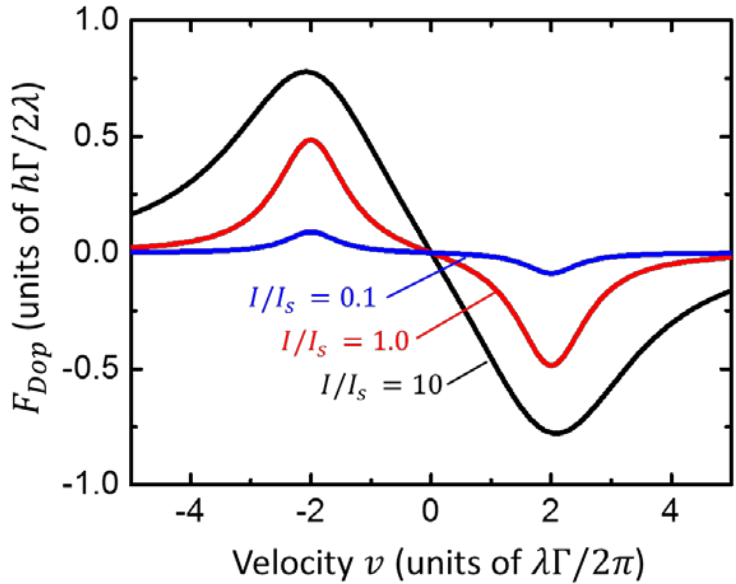

FIG. 5. Doppler cooling force $F_{Dop}$ (Eq. 176) as a function of velocity $v$ for a detuning of $\Delta = 2\Gamma$ at $I/I_s = 0.1$, 1.0, and 10.

The consequence of a limited capture range is that it is sometimes difficult to capture and cool a large number of atoms from a thermal atomic beam, where velocities are usually hundreds of meters per second, without first slowing them down by a significant amount. This problem can be addressed by using the light force in a more direct way to slow the thermal atomic beam. In a typical configuration, a resonant laser beam is counter-propagated with the atomic beam, scattering photons from the atoms to decelerate them. For this to be effective, it is necessary to keep the atoms in resonance with the laser beam as they decelerate, despite their changing Doppler shift. One common approach to maintaining resonance is to use a carefully tailored spatially varying magnetic field to produce a varying Zeeman shift along the deceleration path that exactly counteracts the Doppler shift. In this technique, known as Zeeman slowing,[116] decelerations as large as $10^6$ m/s can be achieved, realizing a continuous beam of slow atoms over a typical distance of 0.5 m to 1 m. Another approach to keeping the atoms in resonance is to use a frequency-chirped laser beam.[117] With this method the construction of a bulky tapered magnetic field solenoid is avoided; however, the cold atoms are necessarily produced in a train of pulses, since the detuning needs to be reset periodically.

Laser cooling is capable of reducing an ensemble of atoms to extraordinarily low temperatures. With Doppler cooling, the lower limit to the achievable temperature is a result of the balance between the cooling process and the heating which arises from the random emission of photons. This limit is solely governed by the spontaneous emission rate for the transition, and is given by $T_D = \hbar\Gamma/2k_B$.[115] For most atomic species, $T_D$ is less than 1 mK,[118] and is typically a few hundred microkelvin.

Although the sub-millikelvin temperatures of Doppler cooling are already very low, even colder temperatures can be attained through more complex laser cooling processes. In fact, the Doppler limit only strictly applies to an ideal two-level atomic system, and in many atoms there are additional energy levels, such as degenerate magnetic sublevels or hyperfine-split levels, which can be exploited to achieve much colder temperatures. A great deal of research has been carried out on ways to do this, and a variety of techniques, generally referred to as sub-Doppler cooling, have been reported. For example, polarization-gradient cooling,[119,120] or Sisyphus cooling,[121] relies on creating a varying potential energy landscape through light-induced energy shifts and then inducing transitions between levels such that atoms repeatedly travel uphill, losing kinetic energy continuously. Another technique, Raman cooling,[122] uses a sequence of Raman and optical pumping pulses to selectively accumulate low-velocity atoms and then re-thermalize the distribution, resulting in a much lower temperature after several cycles. Other techniques, such as "dark-state" cooling,[123] "grey molasses"[124] or cavity cooling[125] use variations on these principles. The temperatures attained with these techniques are generally limited on a scale governed by the so-called recoil temperature, $T_r = h^2/(2k_B\lambda^2 m)$, where $m$ is the mass of the atom. This temperature is associated with the kinetic energy imparted to the atom by the recoil of a single photon. It depends on both the laser cooling wavelength and the atomic mass, and can be very low for heavy atoms; Cs, for example, has $T_r = 0.2$ μK.

*Cold atom traps*

While laser cooling can effectively cool atoms to very low temperatures, it is often necessary to realize some form of trap to contain the cold atoms to prevent them from eventually diffusing out of the cooling region. For atoms with a net magnetic dipole moment, a trap can be realized with a magnetic field having a local minimum, such as a quadrupole field, a so-called Ioffe-Pritchard field, or other configurations.[105] Alternatively, cold atoms can be confined by a purely optical trap, formed at the focus of a very strong, off-resonant laser beam.[126] These traps generally create a conservative potential well in which the cold atoms can be held so that experiments can be performed on them.

Conservative traps provide an ideal container for cold atoms if the goal is to hold the atoms without heating them so they can be cooled further with methods such as evaporative cooling[127] in order to realize, for example, Bose-Einstein condensation. However, if a large population of cold atoms is desired with temperature in the range of the Doppler temperature (or somewhat lower), it is possible to utilize light forces in combination with magnetic field gradients to realize both a velocity-dependent cooling force and a position dependent restoring force in a configuration known as a magneto-optical trap, or MOT.[128]

Because of their simplicity of construction and robust performance, MOTs have become ubiquitous in AMO physics, often providing the first step in collecting cold atoms for further experimentation. A MOT is realized by combining a quadrupole magnetic field, which is zero at the center of the trap and increases in magnitude in all directions, with three pairs of counter-propagating laser beams aligned along three orthogonal directions and aimed at the trap center (see Fig. 6a). The laser beams are circularly polarized, with handedness chosen according to the local direction of the magnetic field on the side of the trap where the beam is incident.

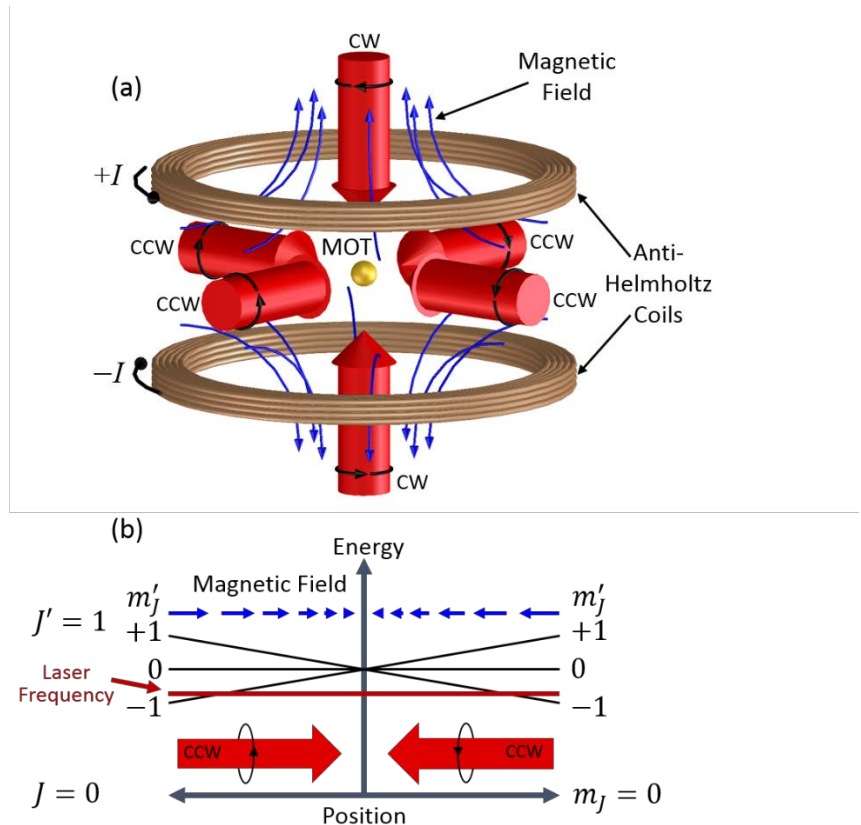

FIG. 6. Magneto-optical trap (MOT). (a) Three mutually orthogonal, counter-propagating pairs of laser beams are aimed at the trap center, defined by the zero of a quadrupole magnetic field created by a pair of coils with opposing currents $\pm I$ (anti-Helmholtz coils). The circularly polarized laser beams have either clockwise (CW) or counter clockwise (CCW) polarization, depending on the sign of the local magnetic field. (b) Energy diagram along one axis of the MOT, showing the Zeeman splitting of the magnetic sublevels of a $J = 0 \rightarrow J = 1$ transition in the presence of a quadrupole magnetic field. With the quantization axis taken along the magnetic field direction, the $m'_J = -1$ level becomes lower in energy and the $m'_J = +1$ level becomes higher. CCW circularly polarized laser light incident from one side of the trap interacts only with $m'_J = -1$ atoms on the incident side and only with $m'_J = +1$ atoms on the other side, where the magnetic field direction is reversed. When the laser is tuned below resonance, the light force on the incident size of the trap is much stronger, creating a net restoring force toward the center of the trap.

While MOTs are quite simple to set up, the principle of operation is somewhat subtle, relying on both laser cooling and spatially dependent light forces that arise from the interplay of a multidirectional magnetic field with laser beams of different polarizations incident from six different directions. In fact, despite detailed studies of MOT behavior,[129] it can be said that not all aspects are fully understood.  For example, with large, high density MOTs, a range of nonlinear, collective effects have been observed which defy simple explanation.[130]

In general, the realization of a MOT requires an atom with a cooling transition in which the excited state has one unit of angular momentum more than the ground state; that is, it requires a $J \rightarrow J + 1$ transition, with $J$ being the ground state angular momentum quantum number.   While $J$ can in principle be any number, in order to understand the basic operation of a MOT it is sufficient to consider the $J = 0$ case.  In such a hypothetical atom, the excited state, with  $J' = 1$, consists of three magnetic sublevels with quantum numbers $m_J' = -1, 0, +1$, as defined with respect to a chosen quantization axis.  In the presence of a magnetic field, these $m_J'$ levels split in energy as a result of the Zeeman effect (see Fig. 6b). If the quantization axis is chosen to be along the magnetic field direction, the $m_J' = +1$ level increases in energy, the $m_J' = 0$ level stays the same, and the  $m_J' = -1$ level decreases in energy.

When near-resonant light is incident on the atom, the magnetic sublevels interact differently with different light polarizations.  Optical selection rules dictate that the $m_J' = +1$ state is exclusively coupled to the ground state by circularly polarized light carrying $+1$ unit of angular momentum (with respect to the quantization axis).  Similarly, the $m_J' = -1$ state is coupled exclusively by circularly polarized light carrying $-1$ unit of angular momentum.  The  $m_J' = 0$ state is coupled only by linearly polarized light. Thus if clockwise (CW) circularly polarized light is incident along the quantization axis, it will interact only with the $m_J' = +1$ state, whereas counter-clockwise (CCW) light will only interact with the $m_J' = -1$ state.

Fig. 6b illustrates an axis of a MOT where the magnetic field points inward, and CCW laser light detuned below resonance is incident from both directions.   On the way into the MOT, the incident light interacts with the $m_J' = -1$ state, scattering photons and creating a light force toward the center of the trap. This force is largest near the outside of the trap and decreases toward the center, because of the Zeeman shift.  After passing through the center of the MOT, this light enters a region where the magnetic field (and the quantization axis) is reversed. Here, the light is still CCW polarized, but traveling opposite to the quantization axis direction, and so carries $+1$ unit of angular momentum from this new perspective.   As a result, the interaction is with the $m_J' = +1$ state, which is Zeeman-shifted well away from the laser frequency, and any light force is weak.  The other laser beam, however, propagating toward the trap center, interacts strongly with the  $m_J' = -1$ state, creating a light force toward the center that increases with distance, just like the first laser beam.  The result of these two counterpropagating laser beams is a net restoring force toward the center of the trap from both directions.  When all three axes of a MOT have such beams, a three-dimensional trap is formed.

The creation of a MOT in practice typically requires only moderate laser intensities of a few tens of watts per square meter and magnetic field gradients in the range of 0.1 T/m, realizable with a few turns of an electromagnetic coil or a collection of small permanent magnets.  Under these conditions it is possible to create a MOT having spatial extent ranging from a few tens of micrometers to a few millimeters.

Unless there are extraordinary circumstances, the MOT population $N$ is generally well described by the simple rate equation

$$\frac{dN}{dt} = R - \frac{N}{\tau}. \tag{18}$$

where $R$ is the load rate, set by the supply of slow atoms within the capture range of the MOT, and $\tau$ is the MOT lifetime, usually determined by the collision rate with background gas molecules in the vacuum chamber. In steady state, Eq. (18) provides a simple expression for the MOT population, $N = R\tau$. With a well-optimized Zeeman slower, load rates of $10^9$ s$^{-1}$ or more can be realized, and lifetimes of several seconds are attainable under ultrahigh vacuum conditions resulting in MOT populations that can reach as high as $10^{10}$ atoms.

The spatial distribution of atoms trapped in a MOT, generally very close to Gaussian for well aligned and balanced laser beams, depends on a number of factors. For low MOT population values, the size of the MOT depends on the strength of the laser cooling transition in the atomic species being cooled, the laser intensity, and the magnitude of magnetic field gradient. As the population increases, however, effects such as internal radiation pressure and light-induced collisional losses result in a limit to the density that can be held in a MOT.[131] A consequence of this is that the MOT grows in size as the load rate and population are increased. A MOT that is only a few hundred micrometers in diameter with low atom population number can easily become a few millimeters in size at a high population number. While the precise density limit depends on the specific atom being cooled and the MOT configuration, densities reported in the literature so far have not exceeded a few times $10^{17}$ per cubic meter, suggesting this is a good rule of thumb for the maximum attainable density in a MOT.

Over the past few decades of experiments using MOTs, a number of variations on the basic geometry shown in Fig. 6a have been developed. For example, a "dark-spot" MOT,[132] with the light intensity at the very center of the MOT blocked to suppress light-induced collisions, has been shown to achieve higher densities. A tetrahedral MOT[133] has been demonstrated with only four incident laser beams. The mirror MOT[134] and the pyramid MOT[135] realize traps near one or more reflecting surfaces, making use of the polarization-manipulating properties of 45° reflecting mirrors to generate the counter-propagating MOT laser beams with the necessary circular polarizations. The grating MOT[136] is similar, with all the necessary beams generated by diffraction of a single large incident beam from gratings strategically placed on a surface. In addition, two-dimensional MOTs have been used as sources of slow atoms[137] or to compress atomic beams.[138]

Because of the need to be able to tune a laser with relatively high precision very close to an atomic resonance, the practicality of laser cooling and trapping for a given atomic species depends on there being an appropriate atomic transition in the species to be cooled. The appropriateness of a transition in turn depends on the wavelength, which must match an available narrowband, tunable laser, and the transition rate, which generally should be in the range of $10^7$ s$^{-1}$ or greater. For magneto-optical trapping and some of the more sophisticated cooling techniques, the transition should also be of the $J \rightarrow J + 1$ type.

Another important characteristic of the transition is that it be "closed" – that is, atoms being excited by the laser must decay only back to the ground state, so they can be re-excited multiple times and thus scatter enough photons to be effectively cooled. In reality, an absolutely closed, two-level atomic transition does not exist, since there are always nearby energy levels to which transitions can happen with greater or lesser probability. In practice, a very nearly closed transition can be realized between hyperfine levels in an alkali metal $D_2$-line, for example between the Cs $6^2S_{1/2}(F=4)$ and $6^2P_{3/2}(F'=5)$ levels, using circularly polarized light. In this case optical selection rules and spectral selectivity combine to optically pump the atoms such that they transition exclusively between the $M=+4$ magnetic sublevel of the ground state and the $M'=+5$ magnetic sublevel of the excited state. Even here, however, the nearby $6^2P_{3/2}(F'=4)$ excited state, separated by 251 MHz, has a small probability of being excited by the cooling laser. As a result, atomic population can leak into the other hyperfine ground state ($F=3$), where it no longer interacts with the laser. In atoms other than alkalis, similar leaks can occur when a metastable state exists at an energy intermediate to the ground state and the excited state. Even a very weak, nominally forbidden, transition to such a metastable level can be problematic if the lifetime is short compared with other timescales in the cooling process, such as the MOT lifetime.

In the face of such leaks, it is often possible to realize a closed system by introducing one or more "repump" laser beams tuned from the level where atomic population is accumulating to an excited state, from where it can decay back to the original ground state. In the case of alkalis, this can take the form of a beam split off from the original laser and frequency-shifted by an acousto-optical or electro-optical modulator to match the transition frequency of the other hyperfine state. It can also consist of one or more separate laser systems, as is done for example in Cr laser cooling.[139] Use of such repumping beams can sometimes add to the complexity of a laser cooling setup.  However, in many cases, the optical leak is weak enough to be ignored, or perhaps repumped with a simple, low-intensity beam flooding the cooling region.

FIG. 7.  Periodic table of the elements, showing the atomic species in which laser cooling has been demonstrated to date.  Note the noble gasses are typically cooled in a metastable state.

Figure 7 shows a periodic table of the elements that have been laser cooled to date.  As seen in the figure, cooling has been accomplished with 27 elements from all parts of the periodic table.  These elements include both metals and non-metals, and span over two orders of magnitude in atomic mass.  In each case the cooling has become possible because of the existence of a suitable cooling transition, the availability of a narrowband, single-frequency, tunable laser of sufficient power, and perhaps the existence of a suitable repumping scheme.  Because a cooling transition can occur in almost any part of the optical spectrum, and must be excited with a resolution typically in the megahertz regime, the availability of the tunable laser has been the most important factor in laser cooling development.  In early years the laser of choice was almost always a tunable dye laser; more recently, the advent of tunable, single frequency diode lasers with wavelengths that match the Cs and Rb cooling transitions has created great opportunities because of their low cost and relative simplicity.  Today, with an increasing availability of laser diodes and fiber lasers at various wavelengths, plus Ti:sapphire lasers with broad tunability, and efficient frequency doubling systems, the opportunities for laser cooling new elements continue to expand.  Laser cooling has even been extended to include certain classes of molecular species, such as SrF[140] and CaF[141].  The present selection of elements shown in Fig. 7, together with the increasing selection provided by new developments, creates a number of opportunities for realization of cold atom ion sources with a broad range of ionic species.

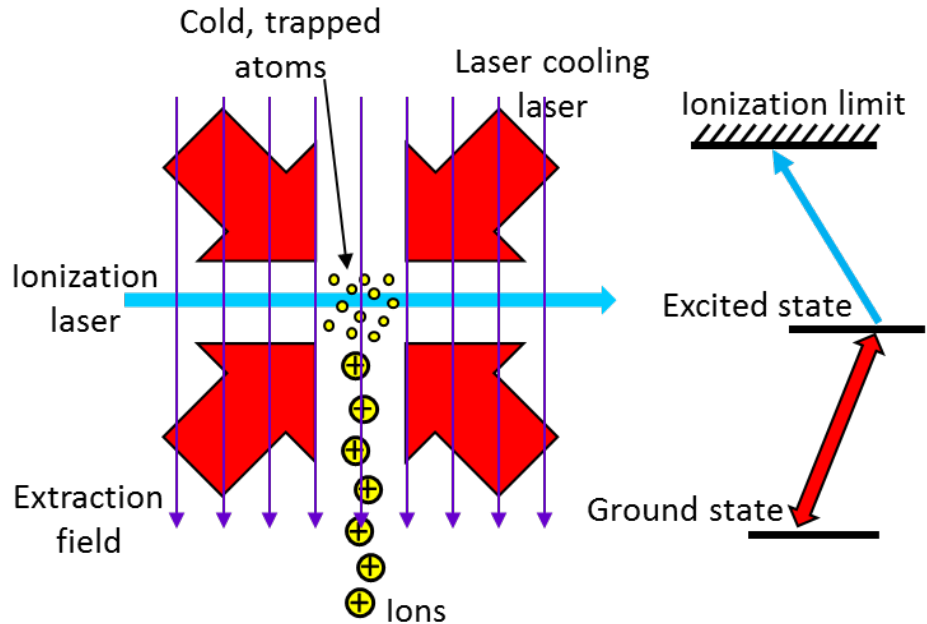

FIG. 8. Magneto-optical trap ion source. A cloud of cold neutral atoms is held in a magneto-optical trap and photoionized in the presence of an electric field. Typically, the ionization laser is tuned from the excited state, which is populated by the laser cooling lasers, to the ionization limit.

## VII.    Magneto-optical trap-based ion source (MOTIS)

We begin our discussion of cold atom ion source realizations with the magneto-optical trap ion source, or MOTIS (Fig. 8). The MOTIS is perhaps the simplest form of cold atom ion source, and has historically been the approach taken by several laboratories as a first demonstration.[142,143] In essence, the source consists of a MOT, ionizing laser light, and a set of electrodes for extracting the ions.

A MOT provides a convenient source of cold atoms for the MOTIS because it is relatively easy to realize with a simple laser and magnetic field configuration, and provides a fairly high density of atoms at temperatures of order 100 μK or less in a relatively small region of space, ranging in diameter from a few tens of micrometers to a few millimeters.

The first step of the ionization process in a MOTIS can be provided by the MOT lasers, as they provide a ready supply of excited state atoms through the laser cooling and trapping process, or via another laser beam tuned to the resonant transition. Ionization is then accomplished by introducing additional laser light to bring atoms from the excited state to the ionization threshold. The wavelength of this light is chosen carefully to just match the transition from the upper level of the cooling transition to the ionization threshold, so that the ions are minimally heated by the recoil of the ionization process.

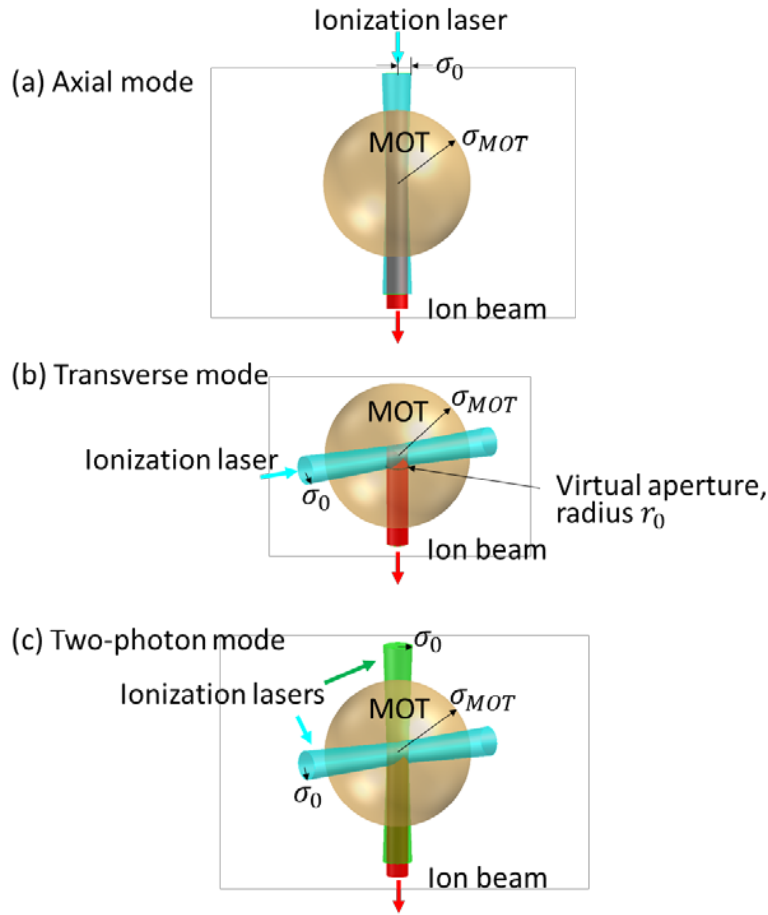

FIG. 9. Ionization modes in a MOTIS. (a) Axial mode, with the ionization laser incident along the ion beam axis; (b) transverse mode, with the ionization laser perpendicular to the ion beam axis and a virtual aperture defined by the extraction optics; (c) two-photon mode, with two crossed laser beams ionizing the atoms in step-wise fashion, defining the ionization volume as the intersection of the two beams.

As shown in Fig. 9, the ionization light can take the form of a single laser beam in one of two orientations, or it can consist of crossed laser beams of two different wavelengths for two-photon ionization. Which type of ionization is chosen depends on the desired source characteristics, and also on the tolerance for complexity in a given set up. With a single beam oriented along the ionization axis (Fig. 9a), referred to as axial mode of operation, a relatively high total current can be obtained because beam current is generated across the entire MOT diameter. However, this comes at the expense of a larger energy distribution, resulting from the electrostatic potential gradient across the MOT, which is necessary for ion extraction. A much narrower energy spread can be obtained with the transverse mode (Fig. 9b). Here, the energy spread is determined by the width of the ionization laser, which can be made quite small. The current in this case is much smaller than in the axial mode, however, as only the central portion of the ionized flux is useful for making a round ion beam. To avoid spurious ions in the beam it is generally necessary in this mode to arrange the extraction optics such that a virtual aperture is created at the source.

With two-photon ionization (Fig. 9c), two laser beams are crossed within the MOT. Ions are only produced where the two laser beams overlap, so the beams can be incident from any direction, as long as their intersection covers the region in the MOT where ion production is desired. This arrangement has the advantage of maintaining a narrow energy spread, while avoiding the unnecessary depletion of the MOT through ionization of atoms not intended for the beam, as occurs in the transverse mode. In this mode, the sum of the two photon energies must just reach the ionization level, either from the ground state of the atom, or from the excited state created by the MOT light. The ionization can proceed via a direct two-photon process, or one laser can be tuned to coincide with an intermediate energy level, allowing resonant two-photon ionization. Whether or not this is practical depends on whether the ionization lasers have sufficient power for the direct two-photon process, or if there is a suitable intermediate level in the atomic species being ionized. One way to accomplish two-photon ionization that works well for pulsed operation is to use light from the MOT laser formed into an auxiliary focused beam as one of the ionization beams. In this case the MOT light is shut off just before the two ionization laser beams are turned on, so that ionization can only occur in the overlap of the two focused ionization beams.

The extraction electrodes in a MOTIS can require a certain amount of engineering, because it can be challenging to get high-precision electrodes providing a uniform extraction field close to the MOT while still permitting the necessary optical access for the MOT and ionization laser beams. With some careful modeling of electric fields and possible use of transparent electrodes, however, this can be accomplished satisfactorily.

As the ions are extracted and formed into a beam, they will necessarily traverse the magnetic field of the MOT, which in principle could be detrimental to the ion-optical properties of the source. While awareness of this is important in designing a MOTIS, generally speaking MOT magnetic fields are modest in strength (field gradients are typically of order 0.1 T/m) and can usually be arranged so that extraction is more or less along field lines, minimizing any perturbations in the ion trajectories. In addition, if pulsed operation is satisfactory, the MOT fields can be turned off during extractions.

*Estimates of MOTIS performance*

To provide some insight into the operating principles of a MOTIS, and to make some estimates of its performance in terms of normalized brightness, total current, and energy spread, we now discuss a theoretical description of the formation of the ion beam from the photoionized cold atoms. As a starting point, we write the spatially-dependent rate of generating ions in the source as

$$P(x,y,z)dxdydz = r_{ion}(x,y,z)n_{ex}(x,y,z)dxdydz, \tag{19}$$

where the $z$-axis is the direction of ion beam propagation, $x$ and $y$ are the transverse coordinates, $r_{ion}(x,y,z)$ is the spatially-dependent ionization rate, and $n_{ex}(x,y,z)$ is the density distribution of atoms in the first excited state – i.e., those available for ionization – as created by the MOT laser light. For single-laser ionization, the ionization rate $r_{ion}(x,y,z)$ is simply proportional to the intensity distribution $I_{laser}(x,y,z)$ of the ionization light, and is given by

$$r_{ion}(x,y,z) = \frac{\sigma_{ion}\lambda}{hc} I_{laser}(x,y,z), \qquad (20)$$

with $\sigma_{ion}$ being the ionization cross section, $\lambda$ the laser wavelength, and $h$ and $c$ Planck's constant and the speed of light, respectively. For two-photon ionization, $r_{ion}(x, y\, z)$ generally takes on a more complex form depending on the type of two-photon ionization being used.[144]

The current density in the beam can be obtained by integrating Eq. (19) along $z$:

$$J(x,y) = e \int P(x,y,z)dz = e \int r_{ion}(x,y,z)n_{ex}(x,y,z)dz, \qquad (21)$$

where $e$ is the fundamental electric charge. From the current density it is straightforward to use Eq. (13), i.e., $B(x,y) = J(x,y)/(\pi k_B T)$, to obtain the normalized brightness of the beam. As long as the ionization process is sufficiently close to threshold, and other heating mechanisms (such as Coulomb interactions) can be ignored, the temperature can be assumed to be that of the cold atoms in the MOT. The total current in the beam is obtained by integrating the current density:

$$I_{TOT} = \int J(x,y)\,dxdy = e \int r_{ion}(x,y,z)n_{ex}(x,y,z)dxdydz. \qquad (22)$$

To obtain the energy spread, we use Eq. (19) to derive an expression for the energy distribution in the ion beam. Unlike conventional sources, the inherent thermal energy spread of a MOTIS is extremely small, considering that a temperature of 100 μK amounts to only 8.6 neV. However, other factors will generally cause the spread to be larger than this. A major contributor to the energy spread is the spatial spread of the ion creation probability in the presence of the electric potential gradient which must be present to create the extraction field. If the extraction field is uniform in the $z$-direction with value $\mathcal{E}_0$, the energy distribution may be written as

$$P(U)dU = \frac{\int r_{ion}\left(x,y,\frac{U-U_0}{\mathcal{E}_0}\right)n_{ex}\left(x,y,\frac{U-U_0}{\mathcal{E}_0}\right)dxdy}{\mathcal{E}_0 \int r_{ion}(x,y,z)n_{ex}(x,y,z)dxdydz}\,dU, \qquad (23)$$

where $U_0$ is the mean energy of the ion beam after acceleration.

Eqs. (19-23) provide a starting point for deriving performance estimates for a MOTIS. Before numerical values can be estimated, however, it is necessary to consider the dynamics of the ionization process. In general, the dynamics in a MOTIS can be quite complex because of the multiple time scales involved. For example, the ionization rate, the MOT load rate, the MOT loss rate (due to background gas and/or inter-atom collisions within the MOT), the cold-atom transport rate, the laser cooling time scale, the thermal relaxation rate, and the ionization and cooling laser time dependences can all play a greater or lesser role, depending on the circumstance. To properly account for time and spatial variation in the source, it is generally necessary to solve a time- and space-dependent partial differential equation incorporating all the ways that ion production can change.

To address the time dependence in a pulsed MOTIS, Debernardi *et al.*[145] present a model which takes into account a number of the important time scales in a MOTIS. For present purposes, we choose two limiting cases: a pulsed mode in which a very fast ionization pulse creates ions on a time scale much faster than any change in MOT population, and a continuous mode, in which a MOT operates at steady state with an ionization laser constantly ionizing atoms at a time-independent rate. We treat these two cases in the following two sections, examining the various modes in which a MOTIS can be operated.

*Pulsed operation*

To obtain pulsed operation estimates, we assume the ionization laser pulse is short, and the time between laser pulses is long, compared with the time scale for MOT population changes. We also presume that we are only interested in the peak, instantaneous performance when the ionization laser pulse first turns on, and so we neglect any depletion of the MOT population caused by the ionization process. Given these assumptions, it is reasonable to write

$$n_{ex}(x,y,z) = \rho_e n_{MOT}(x,y,z), \tag{24}$$

where $n_{MOT}(x,y,z)$ is the unperturbed, steady-state atom density distribution in the MOT without ionization, and $\rho_e$ is the fraction of MOT atoms in the excited state, i.e., those that are accessible to the ionization laser, assumed to be uniform across the ionization region. For axial-mode ionization, we can write the ionization rate at the (assumed instantaneous) onset of the laser pulse as

$$r_{ion} = R_{ion}^0 \exp\left[\frac{-(x^2+y^2)}{2\sigma_0^2}\right], \tag{25}$$

where $R_{ion}^0 = \sigma_{ion}\lambda P_0/(2\pi\sigma_0^2 hc)$, $P_0$ is the peak ionization laser power, $\sigma_0$ is the laser beam size (one standard deviation), and we have ignored any $z$-dependence in the laser intensity, assuming the MOT size is much smaller than the ionization laser's Rayleigh length. Assuming a Gaussian MOT-atom density distribution with standard deviation $\sigma_{MOT}$, we write

$$n_{ex}(x,y,z) = \rho_e n_{max} \exp\left[\frac{-(x^2+y^2+z^2)}{2\sigma_{MOT}^2}\right], \tag{26}$$

where $n_{max}$ is the peak atom density at the center of the MOT. From Eqs. (13) and (21-26) we can now write

$$B_{peak} = \sqrt{\frac{2}{\pi}} \frac{eR_{ion}^0 \rho_e n_{max} \sigma_{MOT}}{k_B T} \qquad \text{(pulsed, axial)} \tag{27}$$

$$I_{peak}^{TOT} = (2\pi)^{3/2} eR_{ion}^0 \rho_e n_{max} \frac{\sigma_0^3 \sigma_{MOT}^2}{\sigma_0^2 + \sigma_{MOT}^2}. \qquad \text{(pulsed, axial)} \tag{28}$$

$$\Delta U_{FWHM} = 2\sqrt{2\ln 2}\, \sigma_{MOT} \mathcal{E}_0 \qquad \text{(pulsed, axial)} \tag{29}$$

To provide a numerical example, we insert nominal values for a Li$^+$ MOTIS. We take $\sigma_{ion} = 10^{-22}$ m$^2$, $\lambda = 350$ nm, $P_0 = 1$ W, $\rho_e = 0.25$, $n_{max} = 4 \times 10^{17}$ m$^{-3}$, $\sigma_{MOT} = 200$ μm, $\sigma_0 = 10$ μm, $T = 300$ μK, and an extraction field of 10 kVm$^{-1}$. The result is a peak normalized brightness of $2.8 \times 10^7$ Am$^{-2}$sr$^{-1}$eV$^{-1}$, a peak total current of 1.4 nA, and a $\Delta U_{FWHM}$ of 4.7 eV.

A similar analysis for the transverse mode yields

$$B_{peak} = \sqrt{\frac{2}{\pi}} \frac{e R_{ion}^0 \rho_e n_{max} \sigma_0 \sigma_{MOT}}{k_B T (\sigma_0^2 + \sigma_{MOT}^2)^{1/2}} \qquad \text{(pulsed, transverse)} \qquad (30)$$

$$I_{peak}^{TOT} = 21.2 \, e \, R_{ion}^0 \rho_e n_{max} \sigma_0^3. \qquad \text{(pulsed, transverse)} \qquad (31)$$

$$\Delta U_{FWHM} = 2\sqrt{2 \ln 2} \, \sigma_0 \mathcal{E}_0 \qquad \text{(pulsed, transverse)} \qquad (32)$$

Note for the total current, a virtual aperture radius of $r_0 = 2\sigma_0$ was assumed, and the factor of 21.2 in Eq. (31) is the numerical value of $4\sqrt{2\pi^3}[I_0(1) + I_1(1)]$, $I_n(x)$ being the modified Bessel function of the first kind. It was also necessary to assume $\sigma_{MOT} \gg \sigma_0$ to obtain an analytical integral. Using the same Li$^+$ MOTIS parameters as above, these expressions result in $B_{peak} = 1.4 \times 10^6$ Am$^{-2}$sr$^{-1}$eV$^{-1}$, $I_{peak}^{TOT} = 95$ pA, and $\Delta U_{FWHM} = 0.24$ eV.

For the pulsed scenario, we do not consider the two-photon ionization scheme because the great variety of ways in which the two photon ionization can be configured leads to too many different specific expressions to include in a general review such as this. Two photon ionization starting from the ground state, starting from the MOT-created excited state, proceeding via a resonant intermediate state, or proceeding via a direct coherent two-photon transition, all result in different expressions that depend in different ways on the intensities of the two laser beams. However, based on the single photon examples provided here, it should not be too difficult for the interested reader to derive an expression for a given specific scenario.

As can be seen from the Li$^+$ MOTIS example, the peak normalized brightness and total current that can be obtained in a pulsed mode represent values that are of significant practical interest. However, for many applications of focused ion beams it is preferable to have a continuous beam. We therefore consider the performance that can be obtained in a continuous mode.

*Continuous operation*

When operated continuously, the MOTIS establishes a steady-state flow of current in the ion beam as neutral atoms become ionized by the ionization laser and are continuously replenished by a supply of fresh atoms loading into the trap. While the dynamics are quite complex in detail, generally speaking there are three distinct processes that can set limits on the total current obtainable from a steady-state MOTIS.

One fundamental limit on current is set by the load rate of atoms into the MOT. If a constant beam current is required, ions cannot be launched into the ion beam any faster than atoms are supplied from the source. While there is no absolute physical limit on the load rate of a MOT, there appears to be a practical limit based on atomic beam and laser cooling principles, at least in what has been demonstrated so far. Essentially, an atomic beam or vapor must be generated, and a sufficient population of these atoms must be slowed to velocities of a few meters per second in order for these to be trapped. Given the state of the art for laser cooling lasers and atomic beam production, load rates up to about $10^{10}$ s$^{-1}$ are what are achievable today. This limit suggests a maximum current in the nanoampere range.

Another practical limit is the ionization rate $r_{ion}$ [Eq. (20)]. Whether or not $r_{ion}$ limits the total current is highly dependent on the specific ion being produced, and also the available ionization laser intensity, since it is proportional to $\sigma_{ion}, \lambda$ and $I_{laser}$. Assuming $r_{ion}$ is the limiting factor, the maximum current that can be produced is of order $e\ r_{ion} n_{ex} V$, where $V$ is the ionization volume. For a continuous Li$^+$ MOTIS with parameters described above, this limit can reach the nanoampere range if the ionization laser power can be as much as 100 mW.

While the load rate and ionization rate can impose overall limits on the total current, there is a third process that can be the limiting factor in a continuous MOTIS even before these practical limits are reached. The continuous production of ions relies on the steady transport of cold, neutral atoms from the surrounding MOT into the ionization region as atoms are ionized and removed.[79] The rate of transport of these atoms is set by the random, thermal motion of cold atoms, which depends only on the MOT density and temperature. As a result, the constraints that exist on the density and temperature of a MOT for a given atomic species become the governing factor in limiting the brightness and total current of a MOTIS.

The transport of cold atoms into an ionization region in a MOT is a complex process with elements of both diffusion and ballistic transport. Considering that a typical elastic cross section is of order $10^{-15}$ m$^2$ for cold atoms at 100 μK,[146] the mean free path at a typical density of $10^{17}$ m$^{-3}$ is several millimeters. Thus the cold atoms in a MOT must be regarded as a very dilute gas, and self-diffusion is not an appropriate description for thermalized motion. Instead, the atoms in a MOT are better approximated as diffusing through a viscous medium created by the optical molasses of the light forces in the MOT.[115] Within this model the trap can be characterized by a spring constant $\kappa$, a damping coefficient $\alpha$, a damping time $\tau_D = \alpha/\kappa$ and a spatial diffusion coefficient $D_x = k_B T/\alpha$.[147] In an equilibrium situation, the MOT size is related to the spring constant by $\sigma_{MOT} = \sqrt{k_B T/\kappa}$, which allows us to write $D_x = \sigma_{MOT}^2/\tau_D$. Generally the spring constant and damping coefficient for a MOT are strongly dependent on the particular atomic species, as well as laser intensity and detuning, and are difficult to calculate accurately from first principles. However, the MOT size and damping time can be measured, and hence the diffusion constant can be derived. For example, measurements on a Li MOT[148] indicate a diffusion coefficient of approximately 7 x10$^{-4}$ m$^2$s$^{-1}$.

Given the diffusion constant, it is in principle possible to calculate the steady-state spatial distribution of cold atoms in a MOT being ionized via a Fokker-Planck equation formulation. However, in a strong ionization limit, which is of most interest because that is where the transport limit is dominant, the density gradient can become very high at the edges of the ionization volume. The Fokker-Planck equation is no longer valid in this region, and the transport is best considered to be ballistic. A way to treat this situation is to assume that ionization occurs with 100 % probability inside a fixed ionization volume, and the transport-limited current is given by the ballistic transport of atoms across the surface of this volume. In this situation, if the ionization volume has a surface area $A_{ion}$, the total current is given by[149]

$$I = \frac{1}{4} e \, n \, \bar{v} A_{ion}, \tag{33}$$

where $n$ is the density of the MOT at the surface of the ionization volume and $\bar{v} = \sqrt{8k_B T / \pi m}$ is the mean thermal velocity of the atoms. Generally, if the ionization region is small and located at the center of the MOT, $n$ can be taken as the peak MOT density. If diffusion is particularly slow in the MOT, however, this density can in principle be reduced by the slow diffusive transport of atoms up to the surface. For a spherical ionization volume of radius $r_0$, with $r_0$ much smaller than the MOT size, it can be shown that the density at the surface is given by

$$n(r_0) = n_0 \left( 1 - \frac{r_0 \bar{v}}{4D_x + r_0 \bar{v}} \right), \tag{34}$$

where $n_0$ is the unperturbed MOT density at the center of the MOT. Using nominal values for a Li$^+$ MOTIS ($D_x = 7 \times 10^{-4} \text{ m}^2\text{s}^{-1}$, $r_0 = 10 \text{ µm}$, and $T = 300 \text{ µK}$) the quantity in braces in Eq. (34) has the value 0.997, indicating that in this case diffusion is not slow enough to significantly reduce the density at the ionization volume surface.

Assuming 100 % ionization and unlimited MOT loading, we can now calculate the peak normalized brightness, total current, and energy spread of a continuous MOTIS in the transport limit. We note that to obtain a simple expression, we must make a further assumption that the ionization occurs uniformly within the ionization volume. In the case of extremely strong ionization, this assumption may break down and the current could arise more from a shell-like region around the edge of the ionization volume. We can also assume that any cold atom that enters the ionization region will eventually be ionized whether it is in the ground or excited state, if we assume it will undergo Rabi oscillations at a rate fast enough to spend enough time in the excited state to be ionized with essentially 100 % probability during its residence time in the ionization region. As a result we do not need to consider the excited state fraction $\rho_e$ as we did for the pulsed case. For the axial mode, we take the ionization volume to be a cylinder of radius $r_0$ oriented along the extraction direction, and the MOT density to be Gaussian as before. We obtain

$$B_{peak} = \frac{1}{\sqrt{2\pi}} \frac{e n_{max} \bar{v} \sigma_{MOT}}{r_0 k_B T} \qquad \text{(continuous, axial)} \tag{35}$$

$$I_{TOT} = \sqrt{\frac{\pi^3}{2}} e n_{max} \bar{v} \sigma_{MOT} r_0 \qquad \text{(continuous, axial)} \tag{36}$$

$$\Delta U_{FWHM} = 2\sqrt{2 \ln 2} \, \sigma_{MOT} \mathcal{E}_0 \qquad \text{(continuous, axial)} \tag{37}$$

Inserting numerical values for our nominal Li$^+$ MOTIS example and taking $r_0 = 10$ μm, we obtain $B_{peak} = 1.9 \times 10^7$ Am$^{-2}$sr$^{-1}$eV$^{-1}$, $I_{TOT} = 480$ pA, and $\Delta U_{FWHM} = 4.7$ eV.

For transverse mode, we consider a cylindrical ionization volume of radius $r_0$ oriented perpendicularly to the extraction direction. As before, we take only current within a circle of radius $r_0$, assuming the extraction optics create a virtual aperture at the ionization region. The result is

$$B_{peak} = \frac{3en_{max}\bar{v}}{4\pi k_B T} \qquad \text{(continuous, transverse)} \qquad (38)$$

$$I_{TOT} = 2en_{max}\bar{v}r_0^2 \qquad \text{(continuous, transverse)} \qquad (39)$$

$$\Delta U_{FWHM} = 1.83 r_0 \mathcal{E}_0 \qquad \text{(continuous, transverse)} \qquad (40)$$

where the factor 1.83 in the energy spread comes from a numerical evaluation of the FWHM for the distribution derived from Eq. (23). Inserting the same numerical values as above, we obtain $B = 5.6 \times 10^5$ Am$^{-2}$sr$^{-1}$eV$^{-1}$ and $I_{TOT} = 12$ pA, and $\Delta U_{FWHM} = 0.18$ eV.

For a two-photon ionization scenario in this continuous, transport-limited case, the details of the ionization process become irrelevant with the assumption of 100 % ionization, and it becomes possible to estimate the performance limits. The ionization region can be modeled as a sphere with radius $r_0$ located at the center of the MOT, and we simply calculate the transport across the surface of this sphere. In this case,

$$B_{peak} = \frac{3en_{max}\bar{v}}{2\pi k_B T} \qquad \text{(continuous, two-photon)} \qquad (41)$$

$$I_{TOT} = \pi en_{max}\bar{v}r_0^2 \qquad \text{(continuous, two-photon)} \qquad (42)$$

$$\Delta U_{FWHM} = \sqrt{2} r_0 \mathcal{E}_0 \qquad \text{(continuous, two-photon)} \qquad (43)$$

and numerical values for the Li$^+$ MOTIS become $B = 1.1 \times 10^6$ Am$^{-2}$sr$^{-1}$eV$^{-1}$, $I_{TOT} = 19$ pA, and $\Delta U_{FWHM} = 0.14$ eV.

Equations (35-43) provide representative expressions for the transport-limited normalized brightness and total current obtainable with a MOTIS in a few ionization geometries. We note, though, that other ionization-laser geometries may be desirable, in which case the expressions would differ. For example, to minimize the energy spread yet keep the total current high, it might be appropriate to use an elliptical beam in the transverse mode, with its short axis aligned along the $z$-axis. For similar reasons a "pancake" geometry could prove useful in the two-photon case.

We also note that Eqs. (35-43) apply only to a MOTIS where pure thermal transport is the only mechanism supplying fresh neutral atoms to the ionization region. It is possible to enhance the transport of atoms, for example with a "pusher" laser beam which uses the light force to induce additional flux into the desired region,[145] or by moving the ionization region around the MOT to sweep up fresh atoms while compensating for the moving source position with synchronously varying ion optics. In this case it may be possible to realize a source with much higher brightness and total current than these equations indicate.

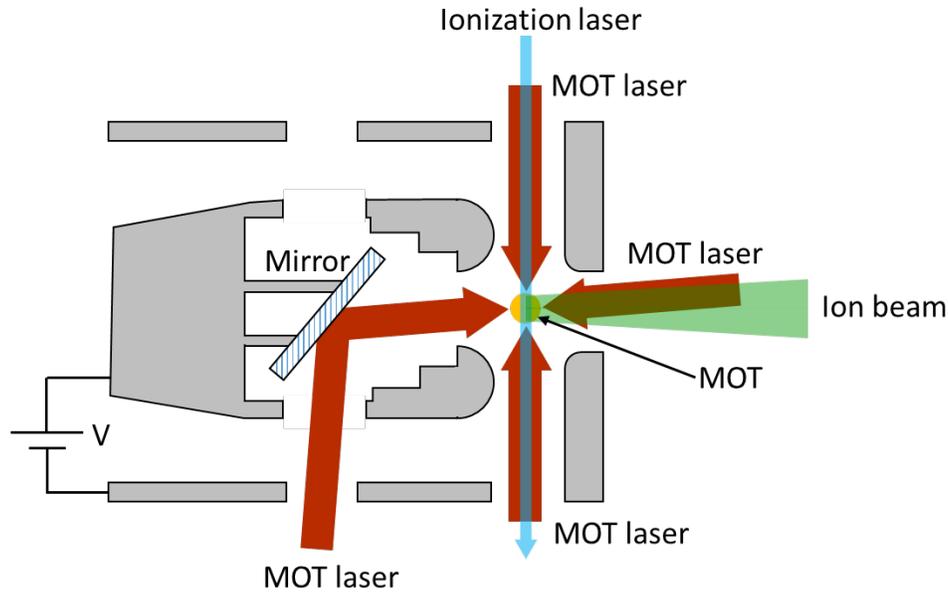

FIG. 10. Rubidium MOT-based ion source. A MOT is formed at the center of a set of extraction electrodes with voltage $V$ applied. One pair of MOT beams is directed nearly collinearly with the ion beam axis by a 45° mirror mounted inside the high voltage electrode. The other MOT beams and the ionization beam are directed radially through holes in the electrodes (adapted from Ref. 150).

From the examples given, the highest brightness and largest total current achievable from a MOTIS come when the MOTIS is operated in pulsed mode with an axial ionization geometry. However, this will not necessarily be the best mode for producing the smallest focal spot FIB. In pulsed mode, the space charge interactions in the high current density of the pulse may reduce the brightness significantly. In axial ionization mode, the energy spread of the ion beam is larger, increasing the chromatic aberrations. These considerations will have to be taken into account when optimizing MOTIS operation for FIB applications.

## VIII.    MOTIS realizations

With the first proposal for using laser cooled atoms to create a bright focused ion beam appearing in 2003,[8] and several similar proposals being published shortly thereafter,[10,142] it soon became evident that practical realization of a magneto-optical trap-based ion source was possible. In this section we review the work that has been done in constructing MOTIS-like sources, describing the principal performance measurements that have been accomplished. Several different approaches have been taken, and we discuss the work categorized by ionic species used.

*Rubidium*

While rubidium does not have a particularly strong motivation as an ionic species for FIB applications, being neither the heaviest nor the lightest alkali, it nevertheless is one of the most convenient atoms to laser cool, with a wavelength of 780 nm, easily accessible by inexpensive diode laser systems. It therefore was a natural choice for initial demonstrations. In some of the earliest work on cold-atom ion sources, Claessens *et al.*[143] described a source of cold electrons and ions based on laser-cooled $^{85}$Rb atoms trapped in a MOT. In this work, the MOT was vapor-cell-loaded, formed in the conventional way by three orthogonal pairs of 780 nm laser beams, and contained about $10^8$ atoms at a maximum density of $6\times10^{15}$ m$^{-3}$. The atoms were ionized by exciting them from the $5p$ $^2$P$_{3/2}$ state to the $44d$ $^2$D Rydberg state with a 6 ns, 1 mJ, 480 nm pulsed dye laser. The Rydberg atoms were then field-ionized by an electric field created by voltage pulses up to 800 V applied to a set of four rods surrounding the MOT. A multichannel plate (MCP) detected extracted ions or electrons, and the resulting pulses of a few picocoulombs lasting a few hundred nanoseconds were observed on a phosphor screen with a charge-coupled device (CCD) camera. Using time-of-flight analysis, an upper limit of (60 ± 30) K was placed on the temperature of the ions in this arrangement.

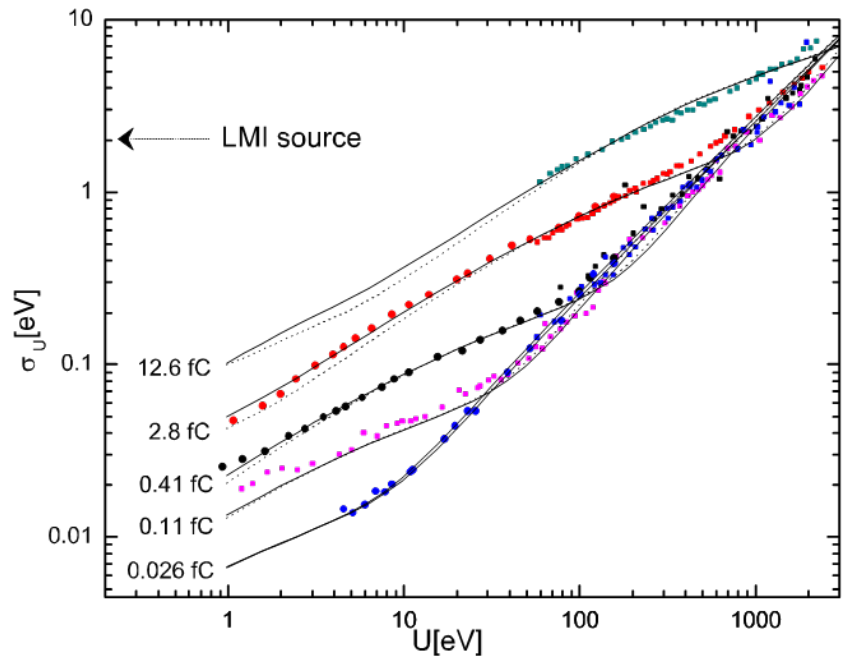

FIG. 11. Energy spread (one standard deviation) $\sigma_U$ of $^{87}$Rb$^+$ pulses from a photoionized MOT as a function of acceleration voltage $U$ for five different pulse charges, showing energy spreads as low as 0.02 eV. Dots are experimental data, solid lines are calculations [reprinted with permission from Ref. 151, M.P. Reijnders, P.A. van Kruisbergen, G. Taban, S.B. van der Geer, P.H.A. Mutsaers, E.J.D. Vredenbregt, and O.J. Luiten, Phys. Rev. Lett. **102**, 034802 (2009)].

Building on this earlier work, Reijnders *et al.*[151] reported on a pulsed ion beam created by photoionizing a vapor-cell Rb MOT held in the center of a specially-formed cylindrically symmetric electrode arrangement.[150] As shown in Fig. 10, two pairs of the MOT laser beams were arranged perpendicular to the ion beam axis, passing through holes in the electrodes. The third pair of MOT beams traveled at a slight angle to the ion beam axis and reflected off a 45° mirror mounted inside the high voltage electrode. The ionization laser was also incident perpendicular to the ion beam axis. In this case the ionization was done with a 2.5 ns pulsed dye laser operating at 480 nm, tuned from the $5p\ ^2P_{3/2}$ state to just above the ionization threshold.

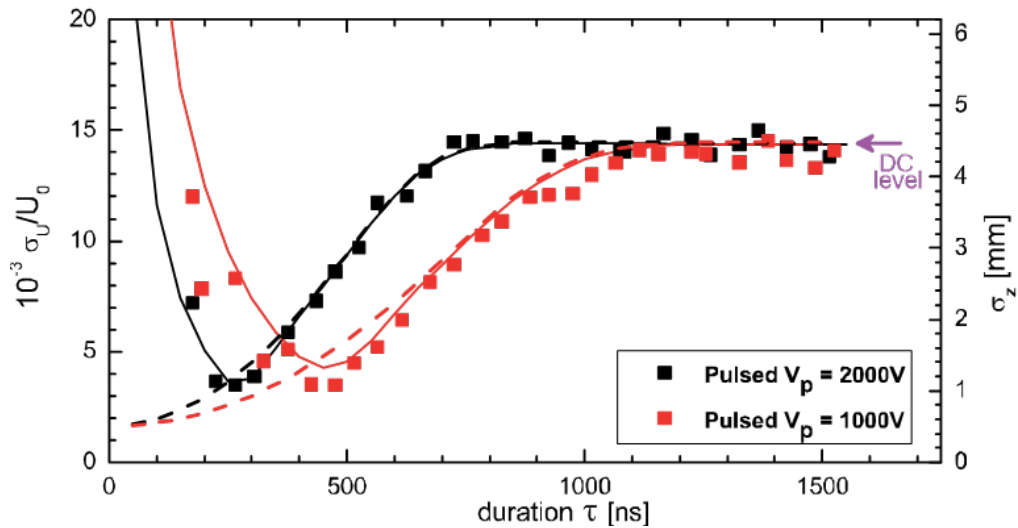

FIG. 12. Relative energy spread $\sigma_U$ as a fraction of the beam energy $U_0$ for $^{87}Rb^+$ pulses extracted from a photoionized MOT. Energy spread is plotted vs duration $\tau$ of the voltage pulse used to extract the ions for two different voltages, showing that a spread significantly smaller than the DC level can be obtained by choosing an appropriate pulse duration. Particle tracking simulations with Coulombic interactions are depicted as solid curves; simulations without Coulombic interactions as dashed curves (from Ref. 153).

Applying DC voltages up to 5 kV to the accelerator electrodes, measurements were made of the energy spread of the Rb ions in the beam by time-of-flight analysis. With pulses of total charge ranging from 0.026 fC to 13 fC, energy spreads were measured as a function of beam energy, and the results are shown in Fig. 11. At the lowest charge values, widths as small as 0.02 eV (one standard deviation) were seen. Such a small energy spread is very attractive for focused ion beam systems where chromatic aberration is a major contributor to spot size. As pulse charge content increased, higher energy spreads were observed. When plotted as a function of beam energy, these measurements showed very good agreement with trajectory calculations that included Coulomb interaction effects.[152]

In subsequent publications,[98,153] these researchers described the use of time-dependent extraction fields to manipulate the energy spread, phase space distribution, and aberrations in a Rb ion beam. The energy spread manipulation is based on exploiting the fact that the energy distribution in a MOTIS is not truly random, but rather correlated with the spatial distribution over which the ion are created. If ions are created in a DC electric field, those that are created on the high potential side of the MOT travel farther in the field and hence have a higher energy than those created on the low potential side. This creates the ordinary energy distribution seen in a MOTIS. If, on the other hand, the electric field is switched off before any of the ions leave the acceleration region, all ions will experience the same force for the same amount of time (assuming a uniform acceleration field), and hence are accelerated to the same final energy. The result is a beam with a reduced energy spread, which can be extremely small, limited only by field inhomogeneities or, ultimately, ion temperature (see Fig. 12). The phase space and aberration manipulation is based on the fact that extraction through a circular hole necessarily imposes focusing on the ion beam, but this focusing can be varied by adding time dependence to the extraction voltage. The result is an ability to sculpt the ion pulse as it is accelerated, focusing or defocusing it, or even modifying the spherical aberration coefficient.

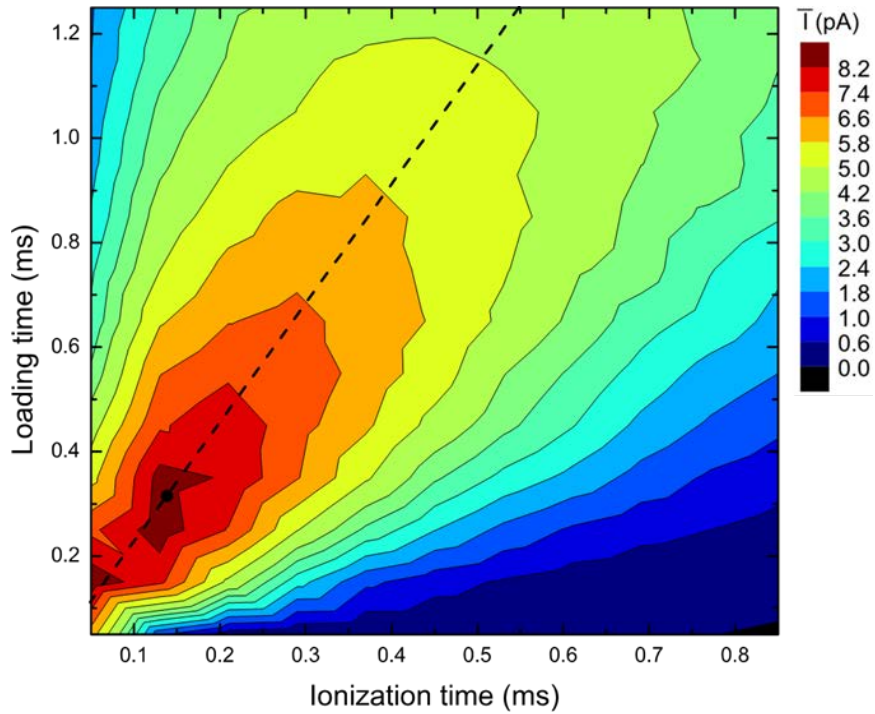

FIG. 13. Average current $\bar{I}$ for $^{87}Rb^+$ pulses extracted from a photoionized MOT vs. ionization time and loading time. The ionization time is the duration of the ionization laser pulse, and the loading time is the delay between pulses during which the MOT is allowed to reload (from Ref. 145).

More recently, some in-depth studies of the ion temperature[154] and optimization of the current[145] have been reported on a Rb MOT-based cold-atom ion source. Using essentially the same apparatus as was used in the work described above, spot sizes measured on a MCP detector were combined with particle-tracking simulations to show an effective source ion temperature of $(3 \pm 2)$ mK. Combining this temperature with source size estimates resulted in a normalized emittance of $1.4 \times 10^{-8}$ m rad $\sqrt{eV}$. While this is a very small emittance, it nevertheless is larger than what might be expected from the MOT temperature (143 μK), even taking into account some additional heating from the ionization laser, which was tuned 0.6 nm above threshold. This additional energy suggests an effective temperature of 390 μK. The authors attribute the difference between this expected value and the measurements to space charge effects and residual distortions present in the beam line.

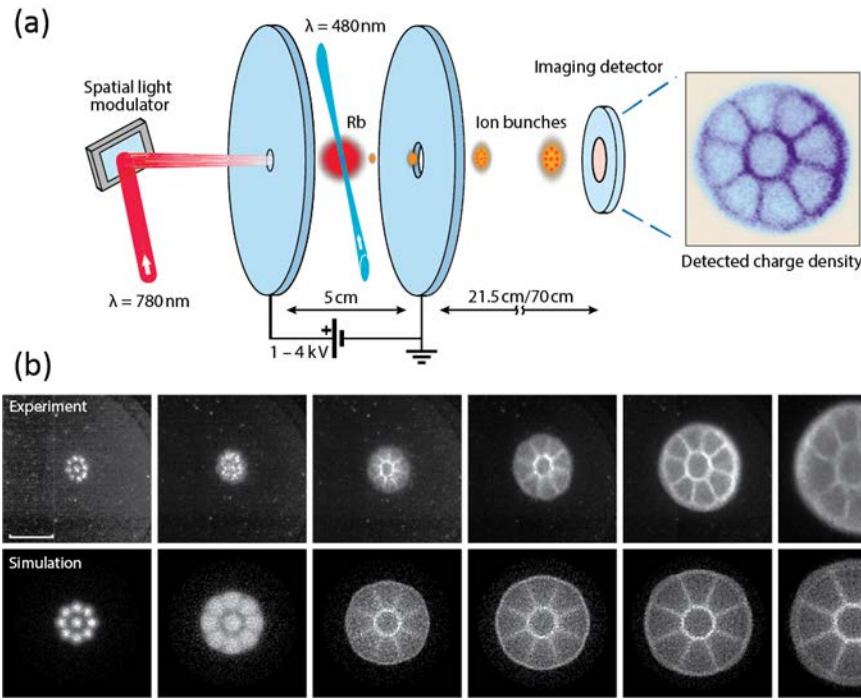

FIG. 14. Observation of space-charge dynamics in a Rb⁺ MOTIS. (a) 780 nm laser light patterned by a spatial light modulator excites Rb atoms in a MOT, which are then ionized by a 480 nm pulsed laser via field ionization of a Rydberg state. (b) Images and simulations of an ion beam patterned in a circular array of nine dots as a function of ion bunch charge. From left to right, bunch charges are ≈0 fC, 2 fC, 8 fC, 10 fC, 13 fC and 19 fC. Scale bar is 5 mm. [Reprinted by permission from Ref. 99, Macmillan Publishers Ltd: Nat. Commun. **5**, 4489 (2014), copyright 2014].

To optimize the current, the average current extracted from the Rb MOTIS was measured as a function of ion pulse length and the delay time between pulses during which the MOT is allowed to reload (Fig. 13). This demonstrates the trade-offs that must be made when operating a MOTIS in pulsed mode. Pulses with the highest peak brightness and highest peak current (such as the numbers given in the previous section) are only produced when the loading time is long, such that the MOT has enough time to attain a maximum density at the start of the ionization pulse. However, long loading times result in a low duty cycle and hence lower average currents, and thus may not be the best choice for FIB applications. Pulses with a peak current as high as 88 pA were generated, limited only by the available ionization laser power (46 mW). While such strong pulses may have utility in certain applications, particle-tracking simulations showed that these high peak-value current pulses suffered from reduced brightness due to stochastic Coulomb heating effects.

In a separate research effort on Rb, Murphy *et al.*[99] made detailed observations of space-charge dynamics in an ion beam produced by photoionizing cold atoms. Using a geometry similar to the chromium MOTIS discussed below, they employed a two-photon ionization approach in which a spatial light modulator was introduced into the excitation laser beam. This allowed the creation of complex two-dimensional patterns in the ionized atoms, enabling detailed studies of beam dynamics. The excitation laser light, provided by a 780 nm laser diode system, created a patterned population of 5P Rb atoms, which were subsequently ionized by illuminating them with 2 mJ to 6 mJ, 480 nm laser pulses of duration 5 ns.  These pulses excited the 5P atoms to an N ≈ 30 Rydberg state, from which they were field-ionized by the extraction electric field.   To investigate the effects of space charge, a pattern of nine dots was created in a circular arrangement, and the ion beam spatial distribution was observed after propagation to an imaging detector (Fig. 14).  Varying the ion bunch charge by varying the ionization laser pulse energy, they observed an evolution from a faithful reproduction of the original dot pattern to complex ring-like structures as the charge in the ion bunches was increased.  The ring-like features, which were well-modelled by Monte Carlo simulations that included Coulomb interaction effects, were attributed to a combination of halos arising from secondarily excited atoms and complex Coulomb interactions between the beamlets in the pattern.

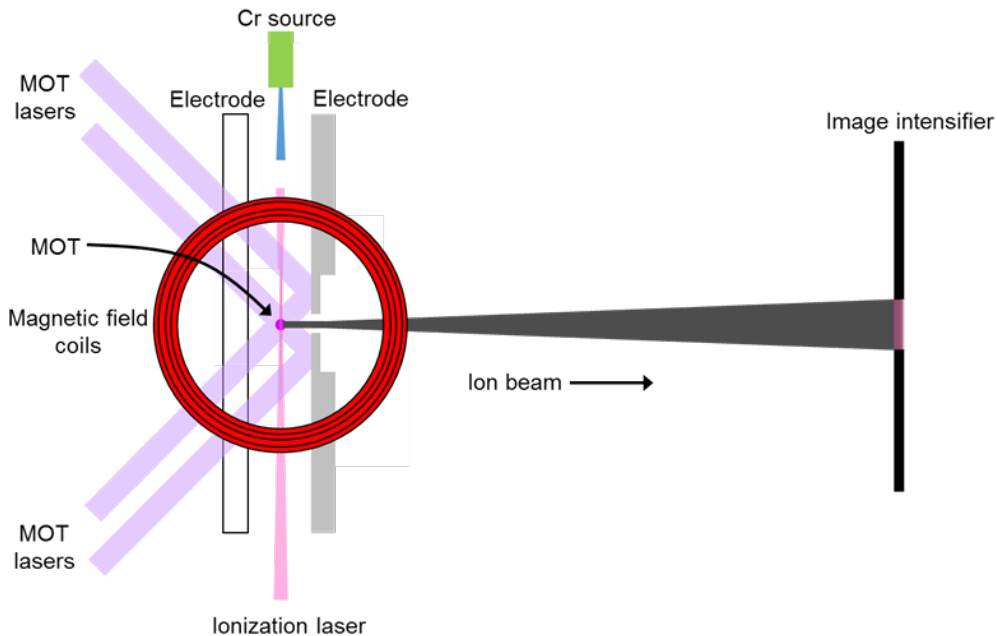

FIG. 15. Cr$^+$ MOTIS.  A MOT is formed between two parallel plate electrodes, one of which is a mirror with a hole for ion extraction, the other being a transparent window with conductive ITO coating.  Two pairs of MOT beams are incident through the window in a "W" configuration; the other pair is perpendicular to the plane of the figure.  Ions are extracted and imaged on a MCP/phosphor/CCD image intensifier detector [adapted with permission from Ref. 156, Nano Lett. **8**, 2844 (2008).  Copyright 2008 American Chemical Society.].

*Chromium*

While chromium, like rubidium, is neither extremely light nor extremely heavy, it nevertheless was an early species in the progression of cold atom ion source demonstrations.  Some motivation for this originated in its use in realizing a MOT-based deterministic, single atom source,[155] but it also has possible uses as a dopant in sapphire (making ruby) and deposition of nanoscale metallic structures.

Beginning with a proposal based on photoionization of laser-cooled chromium,[142,157]  Hanssen *et al.* constructed a source using a folded beam geometry that allowed both uniform extraction field and convenient laser access to the MOT region.[156] In this source, shown in Fig. 15, the extraction field was created by two parallel plates, one of which had a mirrored surface with a hole for extraction of the ions, and the other consisted of a fused silica window coated with the transparent conductor indium tin oxide (ITO). Four of the six MOT laser beams were introduced through the transparent electrode and reflected off the mirrored electrode to intersect at the center in a "W" configuration. The remaining two MOT beams were incident as a counter-propagating pair perpendicular to the ion beam axis. The magnetic field was generated by a pair of coils with opposing currents mounted outside the vacuum chamber. Cr atoms were loaded from a collimated atomic beam source directed at the MOT from the side. Since the primary goal of this study was not to generate as much current as possible, no attempt was made to slow atoms coming from the source in order to enhance MOT loading. Thus only a small fraction of atoms in the low velocity tail of the distribution – those below the MOT capture velocity – were loaded.

In a study aimed at showing that the emittance of the MOTIS can in fact be as low as what would be expected by the MOT temperature and size, a MOT was created with a nominal steady-state atom population of 3300 atoms, a radius of 75 μm (one standard deviation), and a peak density of $(5 \pm 1) \times 10^{14}$ m$^{-3}$. A frequency-doubled Ti:sapphire laser was used to produce the necessary 425 nm light for laser cooling. Ionization laser light at 322 nm was provided by a frequency-doubled ring dye laser, focused to a $1/e^2$ radius of 10 μm, and directed across the MOT in transverse mode.

To obtain a measure of the beam emittance, the ions were extracted by applying a voltage ranging from +50 V to +10 kV to the transparent electrode. The resulting beam was allowed to propagate for approximately 1.2 m before being observed by a MCP/phosphor screen/CCD camera detection scheme. The measurement consisted of observing the width of the ion beam "stripe", which was essentially a projection of the ionization region formed by the ionization laser crossing the MOT, as a function of extraction voltage. To avoid any distortions arising from traversing the MOT magnetic field, the experiment was conducted in a pulsed mode, with the MOT coils being turned off during extraction and propagation of the ion beam.

Because the extraction geometry was a simple pair of plates with a single aperture, it corresponded to a Davisson-Calbick lens[158] with negative focal length that was independent of extraction voltage. Thus any change in beam width as the voltage was varied could be attributed solely to a finite non-zero beam emittance. This can be seen by considering that, with no apertures in the beam, the emittance at the target was necessarily equal to the emittance at the source,

$$(< x^2 >< \alpha_x^2 > - < x\alpha_x >^2)^{1/2} \sqrt{U} = \sigma_x \sqrt{k_B T/2}. \qquad (44)$$

Thus if $U$ is lowered, one expects a corresponding increase in the beam's spatial extent. In fact, the measurements showed no detectable increase in beam width within the measurement uncertainty at extraction voltages as low as 50 V. This result sets an upper limit consistent the emittance of $3.6 \times 10^{-10}$ m rad $\sqrt{eV}$ expected for the measured MOT temperature of $(120 \pm 50)$ μK. Further measurements with higher MOT temperature or shorter ionization wavelength showed clearly detectable increases in the beam width at low energies, consistent with the expected increased emittance.

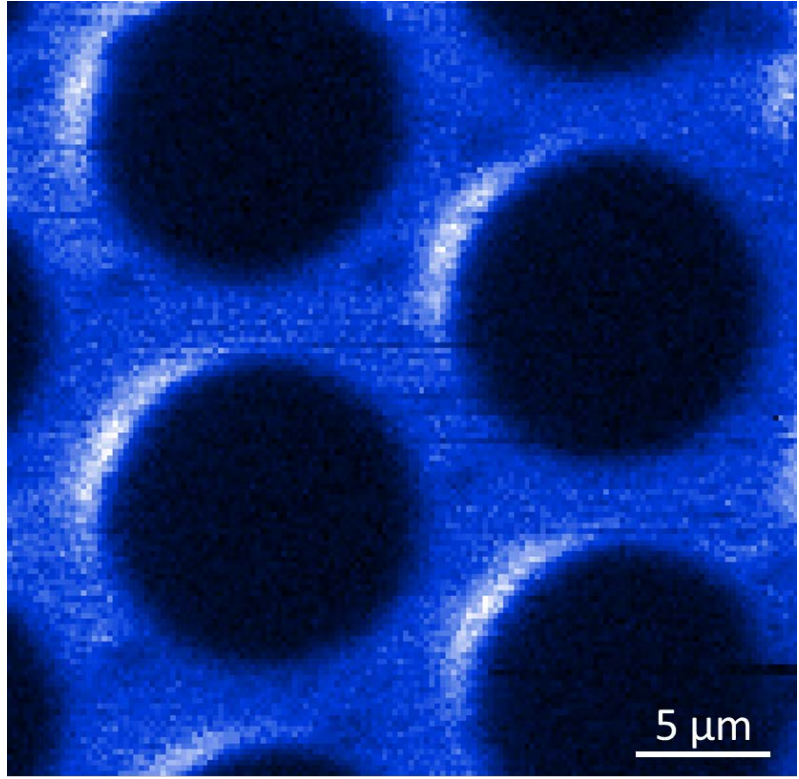

FIG. 16.  Image acquired with Cr$^+$ FIB produced in a MOTIS.  Sample is a MCP surface, with a hexagonal array of 10 μm holes (from Ref. 159).

In a second generation experiment, Steele *et al.*[159] demonstrated a continuous Cr$^+$ MOTIS operating in the axial mode.  In this case the source geometry was the same as it was in the earlier Cr work, with one mirrored and one transparent electrode creating the extraction field.  Instead of using an electromagnetic coil, however, the MOT magnetic field was produced by a pair of ring-shaped NdFeB permanent magnets, considerably simplifying the construction.  Also, a Zeeman slower was used to concentrate the atoms emerging from the Cr source into a low velocity distribution, enhancing the load rate of the MOT.  In addition, the extracted ions were accelerated gradually, using a 265 mm long resistive glass tube.  This acceleration tube provided a way to bring the ions to the desired beam energy of several thousand electron volts, while keeping the electric field in the MOT region low enough to maintain a small energy spread.  It also minimized the focusing of the ions as they were accelerated, creating an essentially parallel beam.  After exiting the resistive tube, the ions passed through a focusing column consisting of a two-axis dipolar deflector and a three element einzel lens.  The focused beam was directed at a sample, and a channel electron multiplier detector collected secondary electrons, realizing a rudimentary focused ion beam microscope.  Continuous beam currents of a few tenths of a picoampere were typically seen with this arrangement.

An image acquired with the Cr$^+$ FIB is shown in Fig. 16. The sample in this case was the surface of the microchannel plate used for beam detection, consisting of a hexagonal array of 10 μm holes in a resistive glass matrix with a metallic coating on the front surface. Fig. 16 represents the first image acquired with a focused beam from a cold-atom ion source. Using an error-function fit to line scans across the edge of a microchannel plate hole, a beam radius (one standard deviation) as small as (205 ± 10) nm was determined at a beam energy of 3 keV. While this spot size is well into the sub-micrometer regime, it is nevertheless approximately a factor of three larger than what was expected from estimates based on the source emittance and ion ray tracing. A likely cause of the discrepancy is the fact that the ion optics were not constructed with sufficient precision, and there was no compensation for astigmatism in the beam. Additionally, there was some evidence from Monte Carlo simulations that Coulomb interactions may have had some effect, and it was also possible that the numerical accuracy of the ion ray tracing was not as high as it should have been. Subsequent images taken using an improved ion optical column with stigmation compensation showed significantly improved resolution.[160]

*Lithium*

The introduction of lithium as an ionic species for cold atom ion sources represents the first purposeful choice based on potential applications. As the lightest of the alkalis, lithium, with atomic weight 7, is heavier only than hydrogen and helium. Thus it is a good choice for ion microscopy, where minimal ion damage to the substrate is desired. In addition, its interactions with target materials are quite different from those seen with helium, the other light-ion microscopy alternative. For example, lithium moves interstitially quite readily in a number of materials such a silicon, and usually diffuses rapidly. Also, it has a smaller likelihood of recombining with a target electron, and thus has a generally higher backscattered ion yield. Furthermore, it plays an active role in many important electrochemical processes, in particular rechargeable battery chemistries, making it an interesting species for ion implantation studies.

While lithium is a clear choice from an applications point of view, it has some relatively minor drawbacks from a laser cooling point of view. Most notable is the typical temperature that can be achieved in a Li MOT. While Rb and Cr cool relatively easily to nearly 100 μK, Li MOTs tend to have temperatures significantly higher than the Doppler temperature, which is 142 μK for Li. Unlike in the other alkalis, where sub-Doppler cooling effects help maintain a temperature at or below the Doppler temperature, Li MOTs have higher temperatures because the hyperfine structure in the excited state is not clearly resolved, preventing such phenomena as polarization gradient cooling. A further consequence of this lack of hyperfine structure resolution is a rather severe optical pumping leakage to another ground-state hyperfine level. As a result, Li requires strong repumping laser light at 803.5 MHz detuning from the main laser frequency. These complications notwithstanding, laser cooling of Li has become relatively routine in a number of laboratories. This is due in part to the availability of relatively inexpensive laser-diode tapered-amplifier systems capable of producing up to 500 mW of the necessary 671 nm light, and also to the relative ease of producing frequency-shifted repump laser beams with acousto-optic modulators.

In preparation for developing a Li ion microscope, Steele *et al.* conducted a study of current production in a continuous Li$^+$ MOTIS.[80] In this experiment, the Li MOTIS was constructed in an arrangement similar to the Cr$^+$ MOTIS discussed above. An upper, ITO-coated transparent electrode and a lower, mirrored electrode created the extraction electric field, while the MOT beams were incident in a "W" configuration. Li was loaded from an atomic beam, which was decelerated by a Zeeman slower to increase the loading rate. The MOT magnetic field was produced by two opposing stacks of NdFeB permanent magnets mounted inside the vacuum chamber, and the 350 nm ionization laser light, generated with a frequency-doubled Ti:sapphire laser, was incident in axial mode.

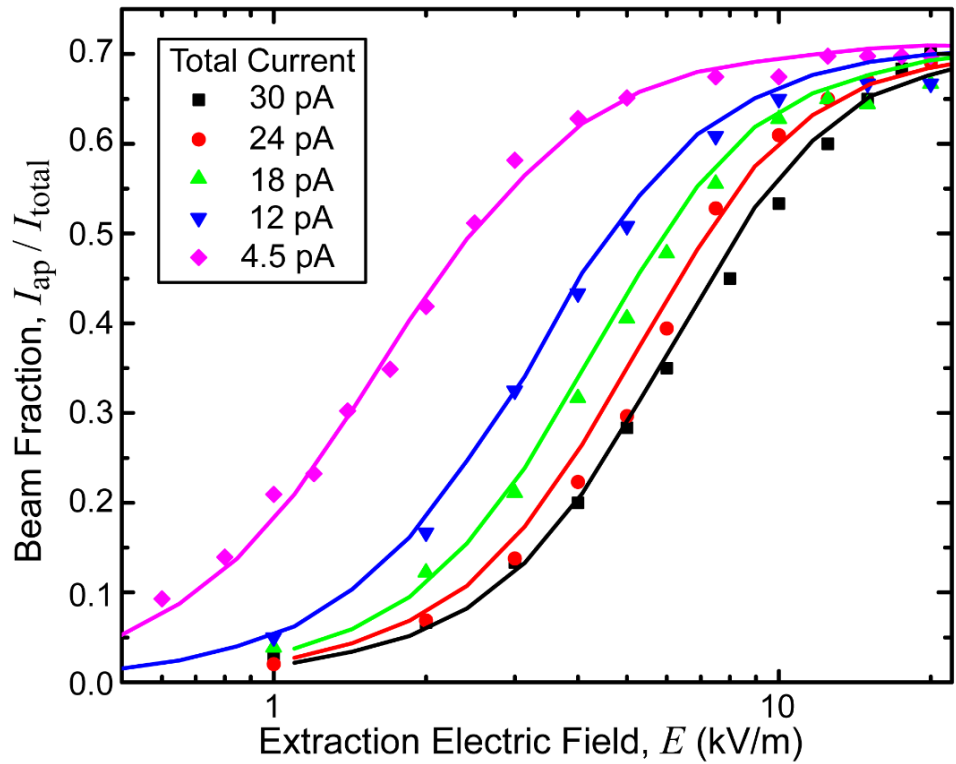

FIG. 17. Coulomb effects in a Li⁺ MOTIS. The fraction of beam current passing a 20 μm aperture is plotted vs extraction electric field for various beam currents. Symbols are experimental data and solid lines are Monte Carlo simulations (from Ref. 80).

The primary goal of the study by Steele *et al.* was to examine the role played by Coulomb interactions in generating an ion beam with a MOTIS. To obtain an experimental measure of these effects, the current that passed through a 20 μm aperture was collected directly after extraction, at a distance of 6 mm from the MOT. With extraction fields of 0.6 kVm⁻¹ to 20 kVm⁻¹, and total currents up to 30 pA, the current fraction that passed through the aperture was measured as a function of extraction field and total current. In the absence of Coulomb effects, one would expect the current fraction to be solely dependent on extraction geometry, and independent of extraction field and beam current. Contrary to this, the experiments showed a marked decrease in transmitted current fraction at low extraction voltages and high currents (see Fig. 17).

To provide deeper understanding of the cause for the reduced transmitted current fraction, Monte Carlo calculations that took into account the individual Coulomb interactions of each of the ions in the beam were carried out. The predictions showed excellent agreement with the experimental results, providing confidence that the Coulomb interaction calculations were capturing the most important aspects of the ion production and extraction.

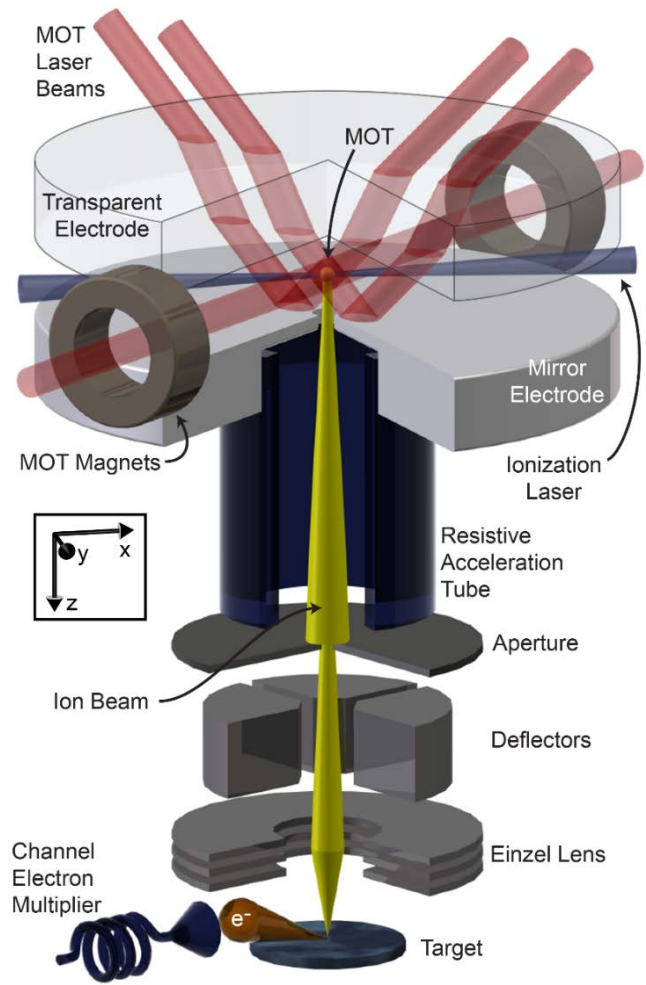

FIG. 18. Li ion microscope based on MOTIS. A MOT is formed between two parallel plate electrodes, one of which is a mirror with a hole for ion extraction, the other being a transparent window with conductive ITO coating. Two pairs of MOT beams are incident through the window in a "W" configuration; the other pair and the ionization laser are incident perpendicular to the ion beam axis. Ions are extracted with a resistive acceleration tube, pass through an aperture, and proceed through a conventional FIB column with deflection, stigmation, and an objective lens (einzel lens). Ions striking a target create secondary elections and backscattered ions, which are detected with a channel electron multiplier (from Ref. 9).

As a result of this study, it became clear that while Coulomb interactions cannot be completely ignored in a MOTIS, there are regimes of parameter space where they can be kept under control to an extent sufficient for robust source operation. For example, the calculations showed that with a 5 pA beam in the MOTIS geometry of this experiment, a relatively modest extraction field of 50 kV m$^{-1}$ (500 V across the extraction plates) was sufficient to make the impact of Coulomb interactions on the source emittance minimal.

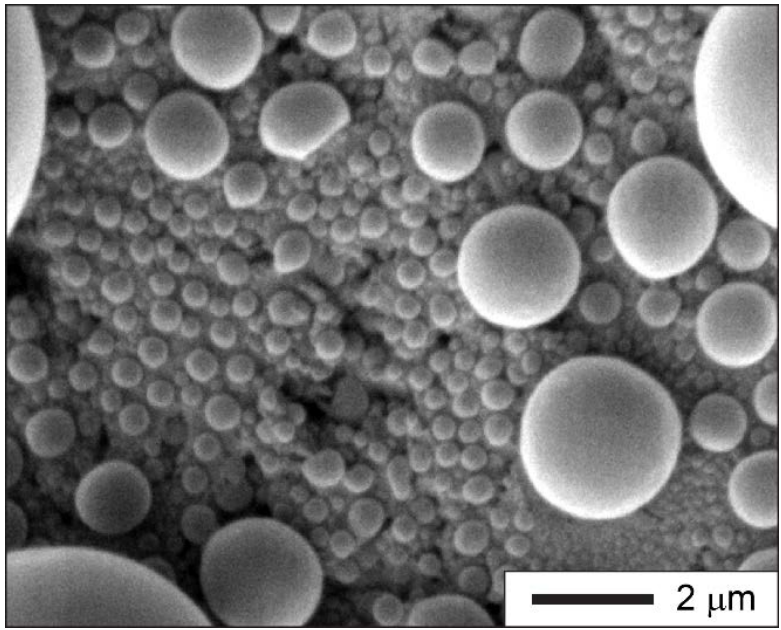

FIG. 19. Secondary electron image of tin spheres on carbon generated by a 2 keV lithium focused ion beam with 1 pA of probe current (from Ref. 9)

Using essentially the same Li⁺ MOTIS, Knuffman *et al.* constructed a Li ion microscope by mounting the source on a commercial FIB platform.[9]  Replacing the Ga⁺ LMIS source with the Li⁺ MOTIS, it was possible to couple the Li ion beam into the entrance of the conventional column, taking advantage of both the high precision construction and the deflection and stigmatic correction capabilities of the commercial system (Fig. 18).  This was accomplished by accelerating the Li ions with a resistive tube, as in the case of the Cr⁺ MOTIS, producing a nearly parallel beam at energies ranging from 500 eV to over 2 keV.   At the point where the Li ion beam entered the column, the first lens, ordinarily used to collimate the diverging Ga ion beam, was disabled, since the ion beam was already collimated.  The Li ion beam then passed through the standard aperture assembly, allowing definition of the beam size in the range 10 μm to several hundred micrometers, and proceeded through the column's quadrupole and octopole deflectors, and eventually the objective lens, to be focused on the sample.

Figure 19 shows an image of tin spheres on a carbon substrate acquired by collecting secondary electrons as the Li ion beam was rastered across the sample.  This image represents the first real demonstration that a quality image can be produced with a cold atom ion source.  Operating in transverse mode, the MOTIS produced a beam current of 1 pA at a beam energy of 2 keV.  Using this beam, a careful study of the focal spot size was carried out by observing line scans across a cleaved silicon knife edge.  This resulted in a $d_{25/75}$ (the 25 % to 75 % rise distance across the knife edge) of (26.7 ± 1) nm, a value fully consistent with the expected beam emittance in combination with small contributions from objective lens aberrations.

*Lithium ion microscopy*

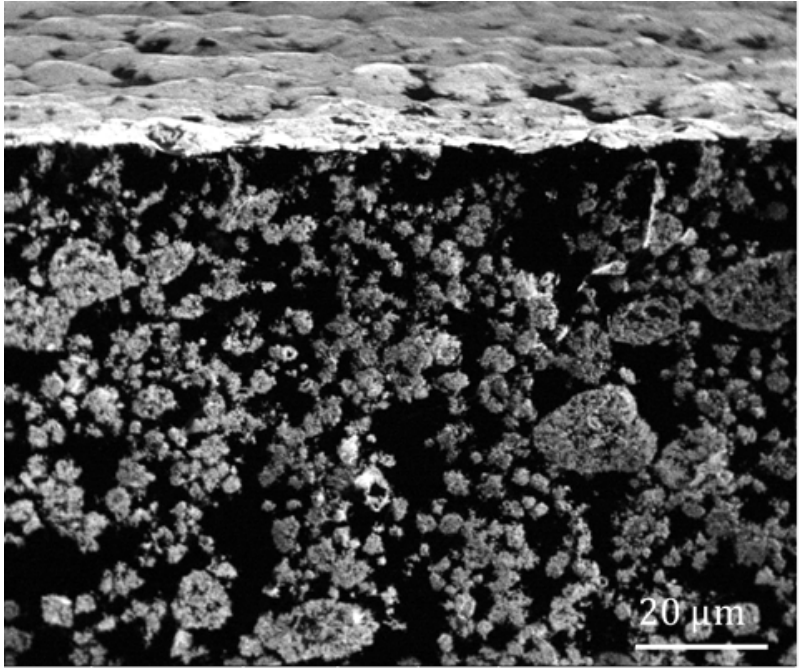

FIG. 20. Li+ MOTIS FIB secondary electron image of nanocomposite consisting of conductive carbon nanotubes embedded in an insulating epoxy resin matrix, showing strong charge-based contrast.  Image is of a cleaved face tilted toward the beam by 8 degrees, and shows a carbon nanotube mat formed on the surface by UV exposures (see Ref. 161).

With the first demonstration of a functional Li ion microscope with decent resolution, it has become possible to begin exploring where a Li+ FIB based on a cold atom ion source might prove advantageous in some of the microscopy and microanalysis applications outlined in Sec. II.  Scanning ion microscopy has attracted renewed attention in recent years since the introduction of the GFIS-based helium ion microscope in 2006.[3]  It seems reasonable that a Li ion microscope could also prove useful, since, like He, Li has a light mass and a low sputtering rate, making it a good choice for microscopy applications due to a smaller likelihood of sample damage during imaging.  There are, however, significant differences between the two types of sources.  While the Li+ MOTIS has a brightness sufficient for nanometer-scale focusing, it nevertheless cannot match the extraordinarily high brightness of a He GFIS.  On the other hand, the Li+ MOTIS is better suited for operation at low beam energies, due to its low energy spread (which scales with beam energy), while the He+ GFIS requires a high extraction voltage and so far has not been shown to operate well below about 10 keV.  This low energy performance of a cold-atom ion source is a unique aspect that can lead to a number of new imaging capabilities.

As discussed in Sec. II, scanning ion microscopy can be performed either by detecting ion-induced secondary electrons or backscattered ions.  Studies with the He+ GFIS have shown[50,162] that ion imaging with secondary electrons has several differences and advantages over imaging with secondary electrons in an SEM, including higher yield, greater surface sensitivity, and positive charging contrast at all beam energies. These advantages arise from significant differences in how electrons and ions interact with a target. For example, electrons have a simple point-charge interaction with the sample, tend to penetrate quite deeply except at the lowest energies, and do not usually cause significant displacement of target atoms. On the other hand, ions have more complex electronic interactions, including the possibility of recombination, have a much shorter penetration depth, and can transfer significant momentum to the target, causing rearrangement and also ejection of sample atoms. While electrons and ions have these significant differences, various light-ion species are not expected to differ much from each other in the processes by which secondary electrons are generated by ion impact. As a result, secondary electron imaging with $Li^+$ should provide the same sort of advantages over a SEM seen with the He ion microscope. As an example, Fig. 20 shows $Li^+$ MOTIS-based microscopy of a carbon nanotube-based composite, where charge-based contrast allows clear differentiation between the conducting carbon nanotube bundles and the insulating epoxy matrix.[161] Because the ion beam consists of positive ions, the insulating areas accumulate a net positive charge and hence do not readily release secondary electrons. This causes them to appear black. On the other hand, the conducting areas allow charge to bleed off and remain essentially neutral, allowing more secondary emission, and show up as a brighter region in the image.

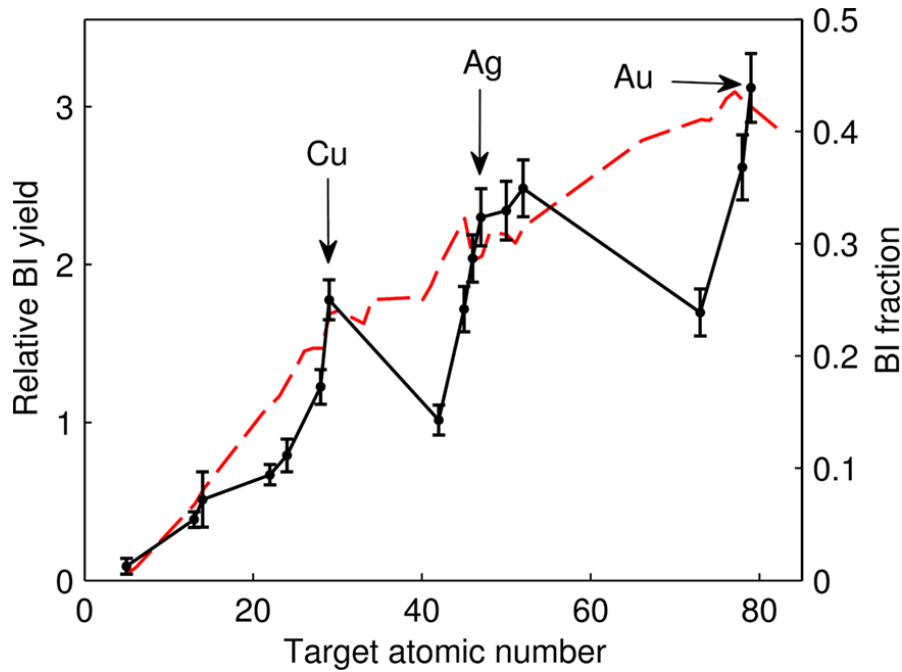

FIG. 21. Fraction of 4 keV Li ions backscattered from pure elemental targets. Symbols connected by a solid line are measurements of individual yields relative to the yield of a stainless steel retainer (relative BI yield), with error bars indicating single standard deviation combined uncertainty. Dashed line is a Monte Carlo calculation of the fraction of ions backscattered (BI fraction) [Reprinted from Ref. 163, "Scanning ion microscopy with low energy lithium ions," by K.A. Twedt, L. Chen, and J.J. McClelland, Ultramicroscopy **142**, 24 (2014), copyright 2014, with permission from Elsevier].

While high quality images such as Fig. 20 can be obtained with a $Li^+$ MOTIS-based FIB, it is generally true that ion imaging with secondary electrons improves as the ion energy is increased, due to a larger SE yield from the primary point of ion impact and fewer extraneous electrons ejected by backscattered ions. As a result, it may become desirable to increase the ion beam energy, in which case the unique low energy capabilities of a cold-atom based Li+ FIB do not provide any special advantage over a He+ GFIS FIB.

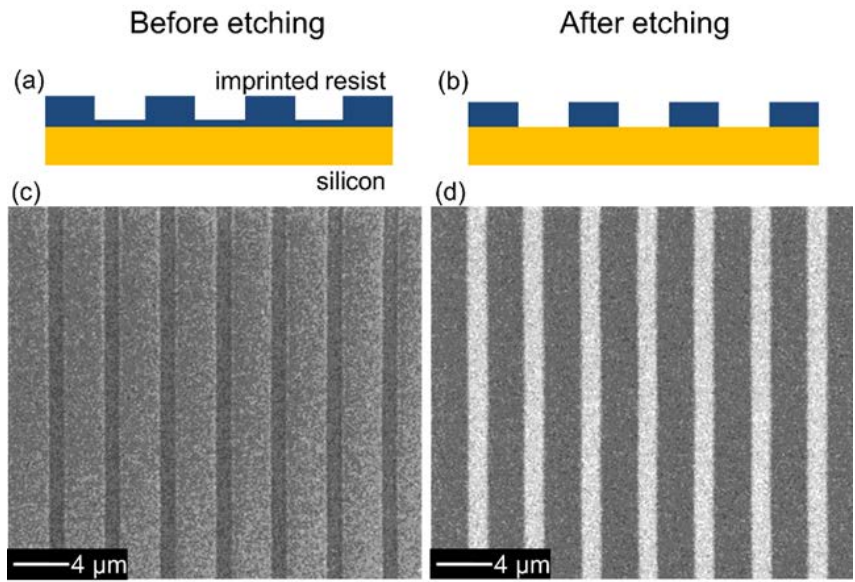

FIG. 22. Backscattered lithium ion images of a nanoimprint lithography grating. (a)-(b) Schematic of imprint and etching process. (c)-(d) Images before and after oxygen plasma etching [Reprinted from Ref. 163, "Scanning ion microscopy with low energy lithium ions," by K.A. Twedt, L. Chen, and J.J. McClelland, Ultramicroscopy **142**, 24 (2014), copyright 2014, with permission from Elsevier].

On the other hand, ion microscopy with backscattered ions improves with lower beam energy, due to an increased backscatter yield and a reduced interaction volume. Furthermore, backscattered ion imaging provides additional information about the sample because different contrast mechanisms are involved. As with backscattered electron imaging in a SEM, the backscattered ion signal primarily gives compositional contrast, since the backscatter yield varies greatly with the atomic number of the target, and is less sensitive to surface charging.

In a study of low energy backscattered ion imaging using the Li+ MOTIS,[163] Twedt *et al.* measured and calculated[164] the relative backscattered ion yield for Li as a function of target atomic number (Fig. 21), and showed several examples of the difference between secondary electron and backscattered ion images. In a particular example of the utility of backscattered ion imaging, they studied the removal of a thin (< 30 nm) residual resist layer during a plasma etching step in a nanoimprint lithography process (Fig. 22). Using the ion beam to image the imprint pattern with backscattered ions, the removal of the residual resist could be confirmed by the stark change in image contrast observed when the relatively high atomic number silicon substrate was exposed by the plasma etch in the trenches next to the low-atomic-number organic resist. The low penetration depth (< 20 nm) of the 2 keV Li ions in the resist material was essential in this case because it prevented the silicon substrate from generating signal through the thin residual resist layer until the layer was fully removed.

While backscattered ion imaging has been investigated with the He⁺ GFIS,[165] it has not seen widespread use, due to the lack of surface sensitivity arising from the higher energy ion beam. In addition, the high resolution advantage of the He⁺ FIB is lost in backscattered ion imaging, since the imaging resolution is set by the interaction volume of the ions in the substrate, which has a size of tens to hundreds of nanometers in the case of He ions with energy in the range 10 keV to 50 keV.

Performing energy analysis on low energy backscattered Li ions from a focused beam could also open up new possibilities in microanalysis by improving the spatial resolution of low energy ion scattering. The Li ion microscope has better spatial resolution than any other light-ion FIB in the energy range of interest in LEIS (1 keV to 10 keV), and is therefore an ideal choice for pushing this surface-sensitive microanalysis technique into the nanometer regime. For LEIS, or for that matter any ion beam analysis technique, the achievable spatial resolution will ultimately be limited in part by the damage done to the sample during analysis. This is because a minimum ion current is generally required in order to obtain a useful signal-to-noise ratio, while the damage scales with the current density at the focus. Concentrating the same amount of current into an ever smaller spot increases the current density and eventually leads to unacceptable damage. This also points to Li as a good choice, since the backscatter yield for Li is higher than for He and other noble gas atoms typically used in LEIS, due to the lower probability of neutralizing within the sample.[61] With a higher backscatter yield, a smaller beam current can be used, and a correspondingly lower current density in the probe can be employed.

Given the performance demonstrated so far by first-generation Li⁺ MOTIS-based ion microscopy, it is clear that there is a role to play for this type of source in the universe of tools available for nanoscale surface microanalysis. While the brightness is not likely to reach the levels seen in a GFIS, the ability to work at low energies can lead to distinct advantages, especially in the area of backscattered ion imaging. As cold atom ion sources continue to develop, it is likely that performance will continue to improve, and new imaging modalities will emerge, leading to a broad new range of microanalysis capabilities.

## IX. Cold atomic beam sources

From the considerations in Section VII, it is apparent that there is a limit to the brightness attainable by a continuously operated magneto-optical trap ion source. Unless one is willing to sacrifice energy spread and use the axial mode, Eqs. (38) and (41) indicate that even with unlimited ionization laser power and atom load rate into the MOT, the normalized brightness is only proportional to the MOT density and inversely proportional to $\sqrt{mk_BT}$, regardless of ionization geometry. Since, for a given atomic species, MOTs have an upper limit to their density and a lower limit to their temperature, the result is an upper limit on the normalized brightness. This limit is entirely due to the reliance on random thermal transport of cold atoms into the ionization region.

If higher brightness is desired, it is possible to create a source in which the atoms do not enter the ionization region randomly, but rather are directed at it using an atomic beam. In this case, the flux of atoms, and consequently the current density of ions, can be increased well beyond the transport limit. This approach has the additional advantage that if the atomic beam travels along the ion beam axis, it only needs to be cooled to very low temperatures in the two transverse dimensions to realize high brightness in the ion beam. Along the axis of the beam, the temperature can be safely ignored because even in a thermally effusive atomic beam, it results in a longitudinal energy spread of only a small fraction of an electron volt, even for atoms with a high evaporation temperature.

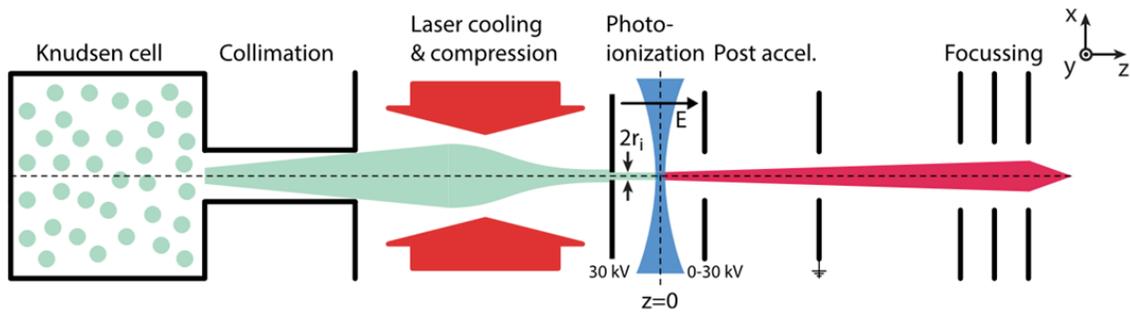

FIG. 23. Cold atom beam source. A thermal atomic beam emerging from a Knudsen cell is mechanically collimated and then transversely laser cooled and compressed. The collimated, intensified atom beam is then photoionized, and the resulting ions are accelerated and focused to form a FIB. (from Ref. 167).

The concept of transversely cooling and then ionizing an atom beam to create a high brightness ion source was in fact one of the earliest proposals for using laser cooling to create bright ion beams.[8] Recently, several authors have expanded on this concept and have provided performance estimates for various configurations of this type of source (Fig. 23). Kime *et al.*,[14] Wouters *et al.*[166] and ten Haaf *et al.*[167] discuss an arrangement where an effusive atomic beam is first transversely cooled and then compressed using a two-dimensional MOT to create a very high flux atomic beam for ionization. The use of a 2D MOT has in fact been in the literature for some time as a way to make a bright neutral atomic beam,[168] and has been shown to be quite effective at compressing and transversely cooling a beam using the same velocity- and spatial-dependent forces present in a MOT, configured to act in two dimensions only. Optimization of the compression is typically achieved by gradually increasing the magnetic field gradient as the atoms move along the compressor axis. Ionization in these arrangements is generally more difficult than in a MOTIS, because the residence time of the atoms in the ionization beam is relatively short, due to the speed of the atoms (usually several hundred meters per second) and the short ionization region length dictated by the desire for a narrow energy spread. Kime *et al.* suggest a Rydberg ionization scheme to circumvent this problem, and ten Haaf *et al.* propose photoionization in an optical buildup cavity, which can enhance the intensity of the ionization laser by a large factor, increasing the ionization rate to the level required for full ionization. The estimates and calculations provided by these authors suggest that a source based on transversely cooling and compressing an atomic beam could in principle, produce a very high performance ion beam with reduced brightness well in excess of $10^7$ Am$^{-2}$sr$^{-1}$eV$^{-1}$.

Currently, realizations of the sources described in Refs. ([14], [166] and [167]) are still under development, although very recently, imaging with a focused Cs$^+$ beam has been demonstrated with this type of source.[169] In an alternate approach, Knuffman *et al.*[170] have constructed and demonstrated a cold atom beam ion source of Cs ions based on a somewhat different concept, with the goal of creating a high brightness source of heavy ions for milling and circuit edit applications (Fig. 24). Rather than using a thermal atomic beam, this source uses a slow atomic beam formed from atoms trapped from a vapor in a 2D MOT. In this arrangement, sometimes referred to as a 2D⁺-MOT,[171] a "pusher" laser beam is incident along the axis of the trap and retroreflected from a mirror with a small hole in it. The pusher and its retroreflection create a velocity-compressed, slow beam which exits through the hole in the mirror. The result is a high intensity atom beam with velocity in the range of 10 m s⁻¹ and relative velocity spread of about 20 %. The low velocity of this atomic beam is advantageous compared with sources based on a thermal beam because it increases the residence time of the atoms in the ionization region, and consequently reduces the requirements for ionization laser intensity. In this source, the atomic beam produced by the 2D⁺-MOT is further intensified using a magneto-optical compressor, followed by a final transverse cooling stage where polarization gradient optical molasses cooling is used to reach as low a transverse temperature as possible. The ionization then proceeds via a two-step process in a pair of crossed laser beams, in which the Cs atoms are first promoted to the first excited state with a focused beam of 852 nm laser light, and then ionized with a focused beam tuned near 508 nm.

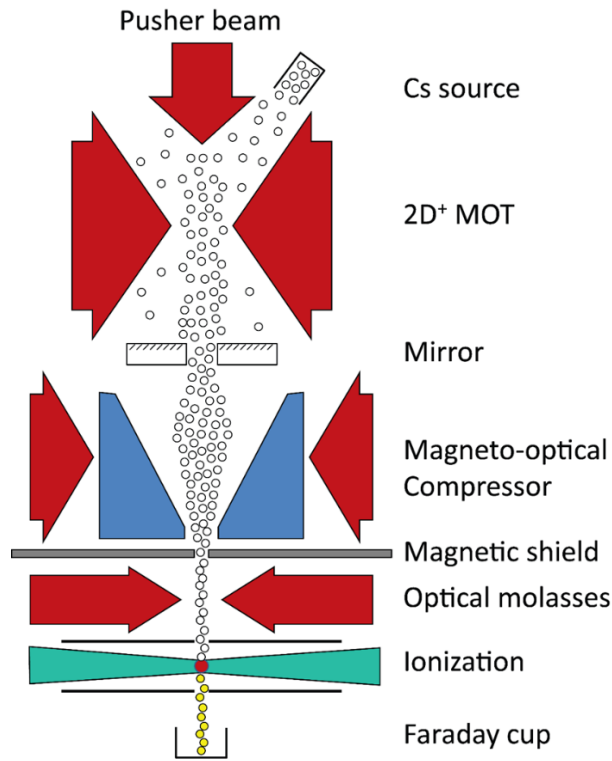

FIG. 24. Cold atom beam source based on a slow atom beam from a 2D⁺-MOT. Atoms are trapped in a 2D-MOT from a vapor and formed into a slow beam with a pusher beam. The beam is transversely cooled and compressed in a magneto-optical compressor, further cooled in a optical molasses stage, and ionized in a crossed-beam arrangement. (from Ref. 170).

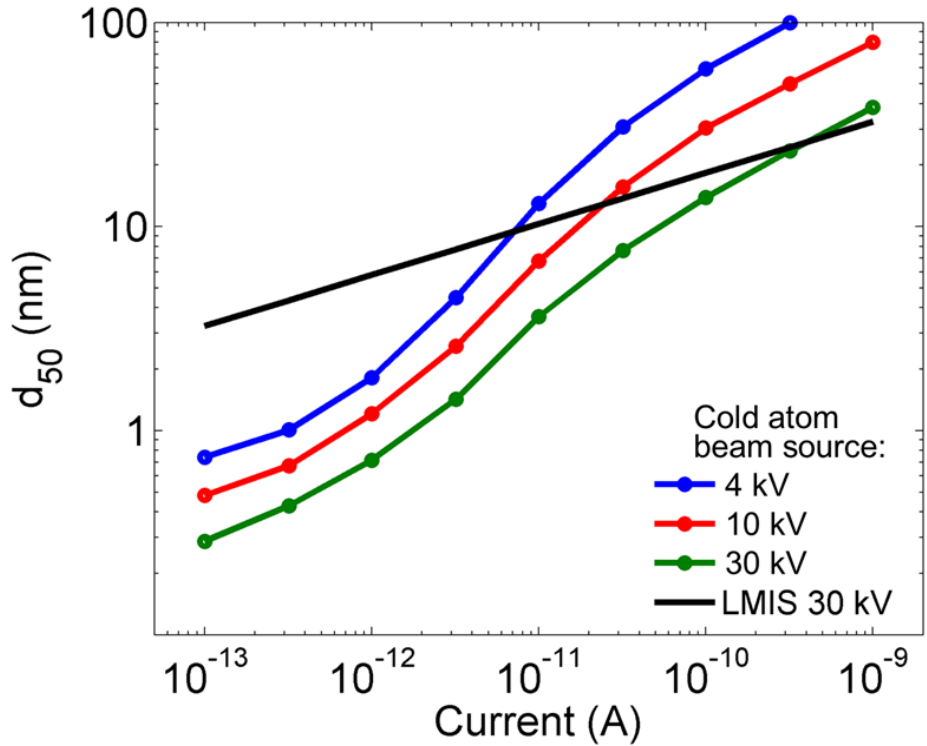

FIG. 25. Projected performance of a Cs$^+$ cold atom beam source for beam energies of 4 keV, 10 keV and 30 keV, together with LMIS data at 30 keV. d$_{50}$ is the diameter into which 50 % of the total beam current falls in the focal plane. Nominal focusing lens aberrations $C_c = 30$ mm and $C_s = 100$ mm are assumed. At each current, an optimum extraction field and corresponding energy spread were chosen to minimize the spot size (from Ref. 172, ISTFA 2014: Proceedings from the 40th International Symposium for Testing and Failure Analysis by ASM International. Reproduced with permission of ASM International in the format Journal/magazine via Copyright Clearance Center.).

To estimate the normalized brightness of a cold atom beam source, it is useful to consider the current density that is generated through the ionization process in the ionization region. We note that in this case the current density is generally a function not only of transverse coordinates $x$ and $y$, but also of the longitudinal coordinate $z$, as the traveling atomic population gradually becomes ionized as it crosses the region, at least until full ionization occur s or the atoms exit the region. If we assume a single velocity $v$ for the atoms and ignore transverse motion, justifiable for a cold source with narrow velocity spread such as a 2D$^+$-MOT, we can write

$$J(x, y, z) = e\Phi_0(x, y)\left(1 - \exp\left[-\frac{1}{v}\int_{-\infty}^{z} r_{ion}(x, y, z')dz'\right]\right), \qquad (45)$$

where $\Phi_0(x, y)$ is the initial atomic beam flux and $r_{ion}(x, y, z)$ is the ionization rate. Eq. (45) can be used together with Eq. (13) to obtain an estimate of the expected normalized brightness by taking $z \to \infty$.

In their first generation realization of a cold atom beam source, Knuffman *et al.* measured a peak current density at the source of 0.16 Am$^{-2}$ while ionizing with an 852 nm laser beam having a one-sigma radius of 2.25 µm and a 508 nm beam having a one-sigma radius of 1.75 µm.  The measured atomic beam radius was 84 µm (one sigma) – considerably larger than the overlap of the two ionization lasers – so the ion beam's transverse spatial dependence was dominated by the spatial dependence of the ionization laser overlap.  This current density, together with a measured transverse temperature of (30 ± 10) µK, suggests a maximum peak normalized brightness of 2×$10^7$ Am$^{-2}$sr$^{-1}$eV$^{-1}$, a value that is quite promising for next generation FIB applications.  A further aspect of this source that adds to its versatility is the fact that if the ionization laser beams are expanded to encompass more of the atomic beam, a much higher current can be obtained, albeit with a lower peak normalized brightness.  Total beam currents in excess of 5 nA were observed in this mode of operation.

While the brightness estimate for this source is encouraging, it does not replace an explicit measurement, because there are a number of effects that can reduce the brightness, in particular Coulomb interactions.  Although an explicit brightness measurement was beyond the scope of the published work, due to the difficulty of measuring the small angular distribution in a nearly parallel high brightness ion beam, it was possible to perform Monte Carlo calculations to estimate the effects of Coulomb interactions.  These calculations were supported by a series of predictions and measurements of the total current as a function of the power in each of the two ionization laser beams.  Generally speaking, statistical Coulomb interactions tend to introduce large transverse velocities for a small fraction of the ions, creating a distribution with long tails.  The standard deviation of such a distribution does not give a very useful measure of the emittance of the beam because it is heavily influenced by a relatively few ions with high transverse velocities.  To account for this, the effect of the interactions was investigated by calculating the brightness for sub-populations of the ion beam with transverse velocities less than some sliding value.  The sub-population brightness was then presented as a function of the fraction of ions included in the calculation.  The results showed that a brightness in excess of $10^7$ Am$^{-2}$sr$^{-1}$eV$^{-1}$ could be expected in a 1 pA beam, including 90 % of the ions.

Using these brightness estimates it was then possible to predict spot sizes for a Cs cold atom beam source, based on nominal focusing lens aberrations of $C_c = 30$ mm and $C_s = 100$ mm.[172]  Fig. 25 shows the results of such a calculation, showing that $d_{50}$ beam diameters well below 1 nm can be expected for beam currents in the picoampere range. This level of performance significantly exceeds the capabilities of the LMIS, suggesting that a number of new applications for focused ion beams may become possible with this source.

## X.      Summary and Outlook

In this review, we have attempted to summarize the current state of the art of cold atom-based ion sources.  While there have been a number of key developments in this field, it is important to recognize that, unlike traditional FIB source technology, this field is very young.  So far only a few research groups have been active, and only a few demonstrations of FIB systems based on cold atoms have been described.  Nevertheless, from the studies that have proceeded so far, it should be clear that this new type of source has the potential to open many new avenues of research into different aspects of nanotechnology.

With the demonstration of Li ion microscopy using a Li$^+$ MOTIS, a number of future directions suggest themselves. First and foremost, an improved Li$^+$ MOTIS can and should be constructed. The first generation source, described in Ref. (9) and used in Ref. (163), while adequate for a number of important demonstrations, does not realize the full brightness that could in principle be approached in the high-ionization, transport-limited limit. With increased MOT loading and ionization laser power, it should be possible to get closer to this limit. In addition, as discussed in Section VII, a number of relatively easy-to-implement improvements can be made, such as introducing a pusher beam to increase flux into the ionization region, or sweeping the ionization region through space to collect fresh atoms while coordinating ion optical compensation for this movement. These could possibly increase the brightness beyond the transport-limited value, creating a source with greatly expanded utility.

Once an improved source is realized, a number of applications can be contemplated. Of particular interest would be the creation of a high resolution, low energy focused Li ion beam for backscattered ion microscopy. The large range of backscattered ion yields as a function of target element suggests the possibility of a new, high resolution elemental mapping capability. Furthermore, if the ion beam is pulsed (easy to do because the ionization is controlled by a laser), time-of-flight energy analysis can in principle be used to pin down exactly what element did the scattering. Such measurements would most likely be best realized in a system with an ultrahigh vacuum sample chamber, to take advantage of the surface sensitivity of this type of low energy ion scattering.

Still more applications will emerge when the specific interaction of a Li$^+$ ion with the target is exploited. An example of this has just been demonstrated, where a focused Li$^+$ beam was used to image the optical modes of a Si microdisk resonator.[173] Here, the gentle impact of the low-energy light ion beam was used to create a nanoscale, localized, reversible swelling of the Si surface, which led to a detectable shift in the resonance frequency. Correlating the shift magnitude with the beam position enabled mapping of the spatial distribution of the optical mode. Such mode mapping promises to be very helpful in designing and building nanoscale optomechanical devices.

Looking beyond microscopy, a high resolution focused Li ion beam has great potential to enable advances in the area of battery material research. Li$^+$ is the ion of choice for a large number of possible rechargeable battery schemes, and key to improving the technology is developing an understanding of how Li$^+$ behaves in various cathode, anode and electrolyte materials. Most candidate materials have complex nanoscale morphology and structure, and it is believed that Li$^+$ transport is profoundly affected by this. With a focused Li$^+$ beam, it will be possible to selectively inject Li ions with nanometer precision into different regions of the material and follow the subsequent transport, thereby unravelling the role played by the structure. This type of application is especially suited for a cold atom ion source, not only because of the high resolution, potential for low beam energy, and simple temporal control of the beam, but also because other ions of interest to battery technology besides Li$^+$, such as Na$^+$ and Mg$^{++}$, are also amenable to laser cooling in their neutral state. With cold atom ion sources constructed for these ions, a great deal of new battery research will become possible.

In the area of nanofabrication via milling, the introduction of the $Cs^+$ or $Rb^+$ cold atom beam source with brightness over $10^7$ $Am^{-2}sr^{-1}eV^{-1}$ promises to enable major progress. In particular, the prospect of an ability to produce a sub-one-nanometer focus of heavy ions like $Cs^+$ with at least 1 pA of current is very encouraging, considering that $Ga^+$ LMIS milling has been stuck in the 5 nm to 10 nm range with little hope of improvement. This advance will be of exceptional interest to the semiconductor industry, where the ability to perform circuit edit on the next generation of chips is still a great uncertainty. In addition to enabling advanced milling, a nanoscale focused $Cs^+$ beam, such as can be realized with a cold atom beam source, has very interesting possibilities for high resolution SIMS, considering that $Cs^+$ has been the ion of choice for analyzing electronegative elements in this application.[94]

As MOT-based and beam-based sources develop, new ways to exploit the exquisite control afforded by laser-atom interactions will also emerge, creating even brighter or more deterministic beams. For example, it has recently been shown that Rydberg interactions have the possibility of suppressing stochastic Coulomb effects, enhancing the ion beam's brightness.[174] These same effects can also result in a quasi-deterministic ion beam from a purely stochastic cloud of atoms.[175] A single "ion on demand" can also be generated using feedback control of loading in a MOT.[155] Such new ways to control an ion beam at the single ion level will open new opportunities for applications like the realization of quantum devices or the controlled doping of materials.

Given this assortment of new capabilities and potential applications, and many more that most certainly will be emerging, there is great motivation to invest research efforts in developing even better cold atom-based ion sources. In the short time over which these sources have been investigated so far, several viable approaches have been proposed and demonstrated. It is fair to say that these only scratch the surface in their use of laser cooling techniques, making use of basic MOT and transverse cooling technology. In the three decades that laser cooling has been evolving since its inception in the 1980s, a great many more sophisticated and varied techniques have been developed. As cold atom ion sources evolve, there will be numerous opportunities for taking advantage of this vast resource of exquisite, high precision methods for controlling the motion of atoms. The enterprise of nanotechnology will undoubtedly see major advances as a result of these developments.

## Acknowledgements

The authors wish to thank M. Jacka, J. Hanssen, D. Narum, J. Orloff, D. Stewart, G. Schwind, M. Maazouz, D. Tuggle, W. Parker, C. Rue, M. Scheinfein, J. Melngailis, E. Vredenbregt, O. Luiten, D. Comparat, R. Scholten and M. Stiles for useful discussions over the years as this field has evolved. KAT acknowledges support under the Cooperative Research Agreement between the University of Maryland and the National Institute of Standards and Technology Center for Nanoscale Science and Technology, Award 70NANB10H193, through the University of Maryland. AVS, BK, and ADS acknowledge support from the National Science Foundation under Grant No. 1353447.